\documentclass[aps,prx,reprint,showpacs,superscriptaddress,longbibliography]{revtex4-2}
\usepackage{mathtools,amssymb,graphicx,units}
\usepackage[usenames,dvipsnames]{color}
\usepackage[plainpages=false,pdfpagelabels,colorlinks=true,linkcolor=blue,urlcolor=magenta,citecolor=magenta,pdftitle={Title},pdfauthor={},pdfdisplaydoctitle=true,pdfduplex=DuplexFlipLongEdge]{hyperref}
\usepackage{subfigure}
\usepackage{grffile}
\usepackage{bm}
\usepackage[version=4]{mhchem}
\usepackage{color}
\usepackage{hyperref}
\usepackage{bibentry}

\usepackage{soul} 

\newcommand\scalemath[2]{\scalebox{#1}{\mbox{\ensuremath{\displaystyle #2}}}} 
\makeatother

\begin{document}
\title{Extended high-harmonic spectra through cascade resonance in confined quantum systems}
\author{Xiao Zhang}
\affiliation{Institute for Theoretical Solid State Physics, Leibniz IFW Dresden, Helmholtzstr. 20, 01069 Dresden, Germany}
\affiliation{School of Physical Science and Technology $\&$ Lanzhou Center of Theoretical Physics, Lanzhou University, Lanzhou 730000, China}
\author{Tao Zhu}
\affiliation{School of Physical Science and Technology $\&$ Lanzhou Center of Theoretical Physics, Lanzhou University, Lanzhou 730000, China}
\author{Hongchuan Du}
\affiliation{School of Nuclear Science and Technology, Lanzhou University, Lanzhou 730000, China}
\author{Hong-Gang Luo}
\email{luohg@lzu.edu.cn}
\affiliation{School of Physical Science and Technology $\&$ Lanzhou Center of Theoretical Physics, Lanzhou University, Lanzhou 730000, China}
\affiliation{Beijing Computational Science Research Center, Beijing 100084, China}
\author{Jeroen van den Brink}
\affiliation{Institute for Theoretical Solid State Physics, Leibniz IFW Dresden, Helmholtzstr. 20, 01069 Dresden, Germany}
\affiliation{Dresden Center for Computational Materials Science (DCMS), TU Dresden, 01062 Dresden, Germany}
\affiliation{Institute of Theoretical Physics and W{\"u}rzburg-Dresden Cluster of Excellence ct.qmat, Technische Universit{{\"a}}t Dresden, 01062 Dresden, Germany}
\author{Rajyavardhan Ray}
\email{r.ray@ifw-dresden.de}
\affiliation{Institute for Theoretical Solid State Physics, Leibniz IFW Dresden, Helmholtzstr. 20, 01069 Dresden, Germany}
\affiliation{Dresden Center for Computational Materials Science (DCMS), TU Dresden, 01062 Dresden, Germany}

\date{\today}

\begin{abstract}
The study of high-harmonic generation in confined quantum systems is vital to establishing a
complete physical picture of harmonic generation from atoms and molecules to bulk solids. Based on a
multilevel approach, we demonstrate how intraband resonances significantly influence the harmonic
spectra via charge pumping to the higher subbands and, thus, redefine the cutoff laws. As a proof of
principle, we consider the interaction of graphene nanoribbons, with zigzag as well as armchair
terminations, and resonant fields polarized along the cross-ribbon direction. Here, this effect is
particularly prominent due to many nearly equi-separated energy levels. In such a scenario, a
cascade resonance effect can take place in high-harmonic generation when the field strength is above
a critical threshold, which is completely different from the harmonic generation mechanism of atoms,
molecules and bulk solids. We further discuss the implications not only for other systems in a
nanoribbon geometry, but also systems where only a few subbands (energy levels) meet this
frequency-matching condition by considering a generalized multilevel Hamiltonian. Our study
highlights that cascade resonance bears fundamentally distinct influence on the laws of harmonic
generation, specifically the cutoff laws based on laser duration, field strength, and wavelength,
thus unraveling new insights in solid-state high-harmonic generation.
\end{abstract}

\maketitle

\section{Introduction}
High-harmonic generation (HHG), originally studied in the gas phase for atoms and molecules  \cite{Ferray1988,McPherson87,Lewenstein1994}, has led to the creation of isolated attosecond pulse \cite{Corkum94,Corkum2007,Sansone443,krausz2009} and, thus, vigorously promoting development of ultrafast technology \cite{Kitzlerprl2005,Shafir2009,Chene1501333}. Recently, the study of HHG has been extended to condensed matter systems \cite{Ghimire2010,Vampaprl2014,Vampa2015,Vampaprb2015,Wang2017,You2017,Luu2015,Younc2017,Schubert2014,Yoshikawa736,LiuH2016,You2016,Hohenleutner2015,Langer2017,Ndabashimiye2016,Uzan2020,Osika2017,Lijinbin2018,Lipra2020,Jimenez-Galan2020,Zhangprb2019,Yueprl2020,Yuepra2020,Klemke2019,Tancogne-Dejean2017,Tancogne-Dejeaneaao5207,Lioe2017,Yucchao2018,Yuchao2020}. Since the first experimental observation in 2011 \cite{Ghimire2010}, solid-state HHG has gained immense interest due to potential applications such as in compact ultrafast light sources \cite{Garg2016,Sederberg2020}, as well as a potential tool to probe microscopic properties of matter, like band structures \cite{Vampaprl2015,Lanin17,Chao2018}, valence electron potentials \cite{Lakhotia2020}, Berry curvatures \cite{Luu2018,Banks2019}, and phase transitions \cite{Silva2019,Alexis2020,Silvana2018,shao2021detecting}.

Depending on the intensity and frequency of the applied electric field, together with details of the band structure, even non-perturbative mechanisms can be responsible for the solid-state HHG \cite{Tamaya2016prl}. In this case, solid-HHG is believed to have contributions from dynamical intraband and interband  processes involving a $k$-space motion of Bloch electrons,  typically described by a three-step model \cite{Vampaprl2014,Vampa2015}. Nevertheless, a clear understanding of the underlying mechanism(s) for solid-state HHG is still under debate \cite{Vampaprb2015,chaoyu2019,Tamaya2016prl}.

For a confined quantum system, due to lack of translational symmetry, the spectrum consists of only a discrete set of energy levels instead of bands. It is known that the HHG spectra in quantum dots can be influenced by confinement conditions such as size and/or coupling parameters in coupled quantum dots \cite{Chuquantumdots2010,avchyan2021high}. One of the distinct advantages of confined quantum systems is the tunability of its energy spectra and wavefunctions by tailoring its size \cite{Murray2000,Bera2010}, or by external parameters, such as gate voltage or magnetic field \cite{PhysRevB.42.5166,PhysRevLett.71.613,PhysRevLett.93.217401}.

A rather interesting situation is realized in partially confined systems, such as quasi-one-dimensional (quasi-1D) systems which are confined in one dimension but periodic in other resulting in a series of subbands in the band structure \cite{PhysRevB.73.165319}. Therefore, such systems possess properties of both bulk and finite systems. The most notable example is perhaps the graphene nanoribbon (GNR) \cite{MitsutakaFujita1996,Nakada1996,Novoselov666,Son2006,Motohiko2008,Ezawa2006}. Depending on the polarization of the electric field, different aspects of HHG, corresponding to the bulk and/or the confined quantum systems can be explored. Arguably, the phenomenology of HHG in such a confined quantum system, therefore, forms a bridge between atoms, molecules and bulk solids. A detailed understanding of the underlying mechanism in such quantum systems may not only shed light on the HHG phenomenology in bulk solids but also unravel new phenomenology and HHG response in confined systems. 

Understanding the HHG phenomenology in such systems necessitates going beyond the two-level (equivalent to the commonly employed two-band studies of the bulk solid-state HHG), thereby raising a number of conceptual questions: For a laser field polarized along the confining direction,  what would happen if the laser frequency  $\omega_{0}$ approximately matches the subband gap, i.e., $\omega_{0}\approx \Delta_{\rm sg}$? Does it cause a resonance resembling Rabi flopping in a two-level system  \cite{Rabi1936,Rabi1937}? If many such (approximately) equi-separated subbands are present, does the resonance involves multiple subbands? If so, could this resonant excitation affect the harmonic process remarkably? In addition, Hansen {\em et al}. indicate that the observed cutoff law on HHG transits from atomic to solid-state type with increasing system size in a model 1D chain \cite{Hansen2018}. Does this phenomenon hold even in the presence of resonance?

Here, by considering a multilevel model and solving the time-dependent Schr{\"o}dinger equations, we demonstrate that the subband resonance can lead to remarkable effects in the HHG spectra. We find that this effect is especially prominent in GNRs due to the presence of nearly equi-separated bands, To this end, as a proof of principle, we consider GNRs interacting with a resonant field polarized along the cross-ribbon (confined) direction where it leads to \textit{cascade resonance}, a multilevel phenomena involving almost all the valence and conduction subbands, and an enhanced HHG spectra well beyond the current cutoff laws.

Specifically, the plateaus of the harmonic spectra are broadened significantly when the laser frequency matches the subband gap. The cascade resonance effect causes the excited electrons to gradually accumulate near the Dirac points also in higher-energy subbands (charge pumping), eventually leading to the extended HHG spectra. Subsequently, we establish the conditions for occurrence of cascade resonance in GNRs. Our analysis indicates that the occurrence of cascade resonance requires a threshold field strength in addition to satisfying the frequency-matching condition. In addition, we formulate the dependence of harmonic cutoff in GNRs on laser duration, field strength and wavelength, when the cascade resonance occurs. These dependencies are significantly different from those of HHG in atoms, molecules, and bulk solids currently studied. Finally, we discuss the possibilities for the cascade resonance in other materials by considering a general multilevel model and implications for HHG spectra.

\section{HHG in Graphene Nanoribbons}
\begin{figure}
	\centering
	\includegraphics[width=\columnwidth]{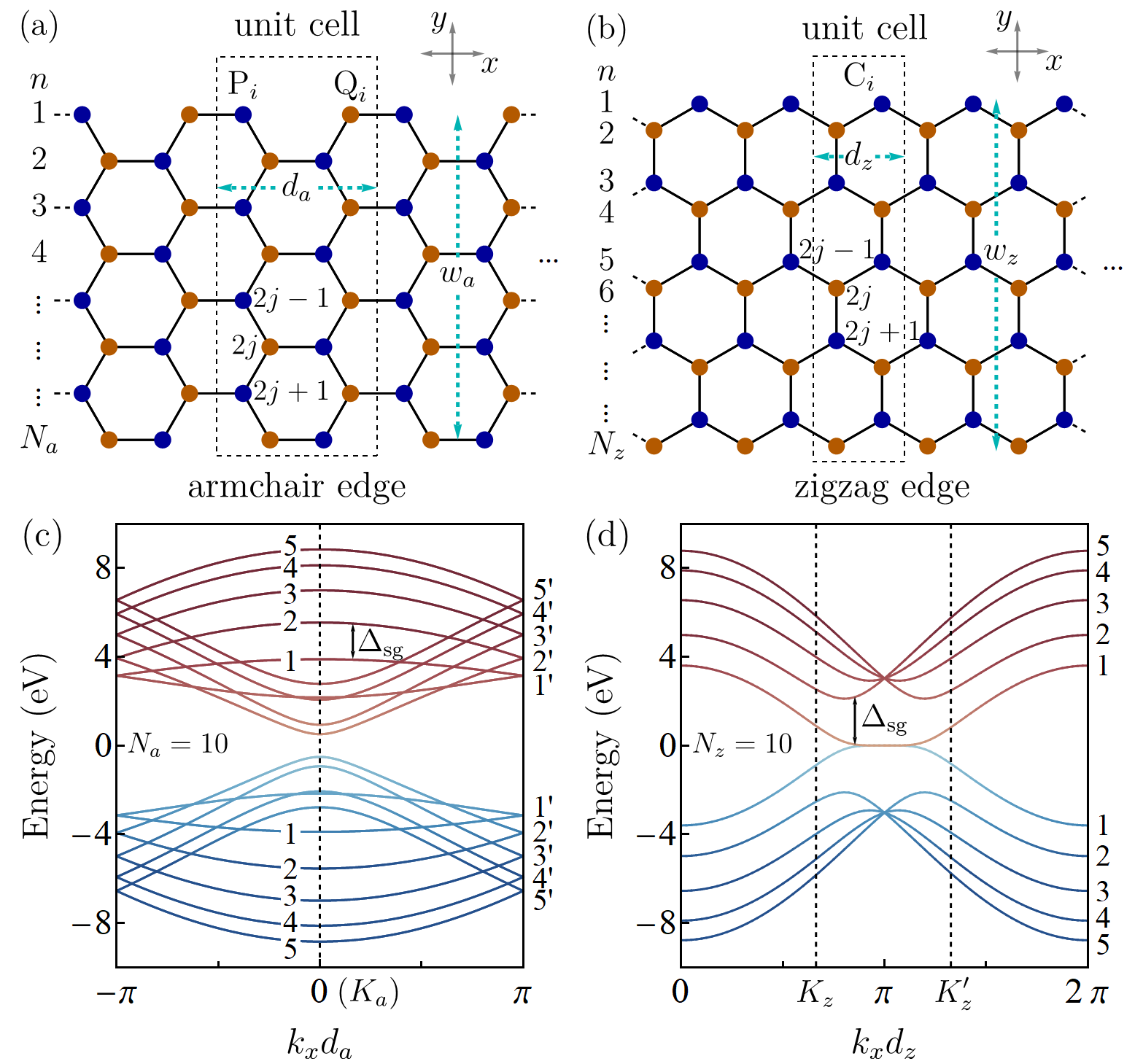}
	\caption{(a)--(b) Structure of graphene nanoribbons with (a) armchair edges and (b) zigzag edges. Blue and orange cycles represent the nearest-neighbor two carbon atoms. (c)--(d) Band structure of (c) 10-AGNR and (d) 10-ZGNR. Red (blue) curves stand for the subbands belonging to the conduction (valence) band. $\Delta_{\rm sg}$ denotes the subband gap. $K_{a}=0$ and $K_{z}=2\pi/3d_{z}$, $4\pi/3d_{z}$ are Dirac points, respectively, for the armchair and zigzag GNRs when the boundary conditions of the transverse ($y$) direction are periodic. $w_{a}$ ($w_{z}$) is the width of armchair (zigzag) nanoribbon. $d_{a}$ ($d_{z}$) is the distance of the armchair (zigzag) unit cell. 
	}
	\label{fig1}
\end{figure}

GNR is a quasi-1D material extending in two directions---ribbon ($x$) and cross-ribbon ($y$) directions \cite{MitsutakaFujita1996,Nakada1996,Novoselov666,Son2006,Motohiko2008,Ezawa2006}. GNR with armchair edges (AGNR) on both sides is classified by the number of dimer lines ($N_{a}$) across the ribbon width as shown in Fig.~\ref{fig1}(a). The unit cell of AGNR consist of two chains P and Q. Likewise, GNR with zigzag shaped edges (ZGNR) on both
sides is classified by the number of the zigzag chains ($N_{z}$) across the ribbon width
[Fig.~\ref{fig1}(b)]. The chain in unit cell of ZGNR is labelled by C. We refer to GNR with $N_{a}$ armchair dimer lines 
as $N_{a}$-AGNR and GNR with $N_{z}/2$ zigzag chains as $N_{z}$-ZGNR. Tight-binding model of GNR is presented in Appendix~\ref{sec:methods}.

Since the ribbon is macroscopically large along the $x$ (longitudinal) direction, continuous band structure can be obtained across the Brillouin zone (BZ) shown in Figs.~\ref{fig1}(c) and \ref{fig1}(d). By contrast, in the $y$ (transverse) direction, transverse confinement gives rise to a discrete set of subbands which are one of the typical features for electronic structure of nanoribbons. Following previous convention, we adopt $J$($s$) as index notation for the ribbon subbands, see Fig~\ref{fig1}(c) and \ref{fig1}(d), where $J = 1, 2,\ldots, N$ is the band number and $s$ is the band type with “$c$” and “$v$” representing conduction and valence bands, respectively. Different from subbands of ZGNR, the subband indices of the AGNR in Fig.~\ref{fig1}(c) are classified into two groups labelled by $\left\{J\right\}$ and $\left\{J'\right\}$ respectively, where subbands $J(s)$ and $J'(s)$ merge at the boundaries of BZ. Atomic units are used throughout the paper unless otherwise indicated.

Recently, there has been a growing interest in studying HHG from GNRs or similar nanoribbons \cite{Cox2017,Wu2020,Christophpra2020,Jurss2021,Drueeke2021}, owing to diverse electronic properties of GNRs, which arise particularly from different edge geometries, \textit{viz.} zigzag and armchair edges. When the laser field is polarized along the ribbon direction, the bulk aspects of GNRs are reflected in the harmonic spectra. For example, the edge states of ZGNR enhance the emitting efficiency of low-order harmonics \cite{Wu2020}; the onsite potential, breaking the mirror symmetry, causes the perpendicular harmonic emission \cite{Jurss2021,Drueeke2021}. It is easy to know that the cutoffs of longitudinal harmonic spectra scale linearly to the field strength and wavelength as shown in Appendix~\ref{ribbonHHG}, which is similar to the HHG in bulk graphene. However, the finite-size effects \cite{McDonald2017prl,Hansen2018} in GNR along the cross-ribbon direction has not yet been fully examined.

In GNRs, the HHG generated by transverse field can arguably be more interesting than generated by longitudinal field. Quantum confinement effects  reflect on HHG when the applied laser field is polarized along cross-ribbon direction. Specifically, the nearly equal-energy spacing subbands play a role and induce a resonance excitation over the subbands. In general, however, optical transitions between two subbands are not always allowed. The optical selection rules for GNRs is a result of the wave function parity factor $(-1)^J$ , where $J$ is the subband index which has been mentioned before. A detailed derivation and discussion of the optical selection rules is outlined in Appendix \ref{sec:selection_rules}.

\subsection{\label{sec:III}Size dependence of the transverse HHG spectra}

\begin{figure}
	\includegraphics[width=\columnwidth]{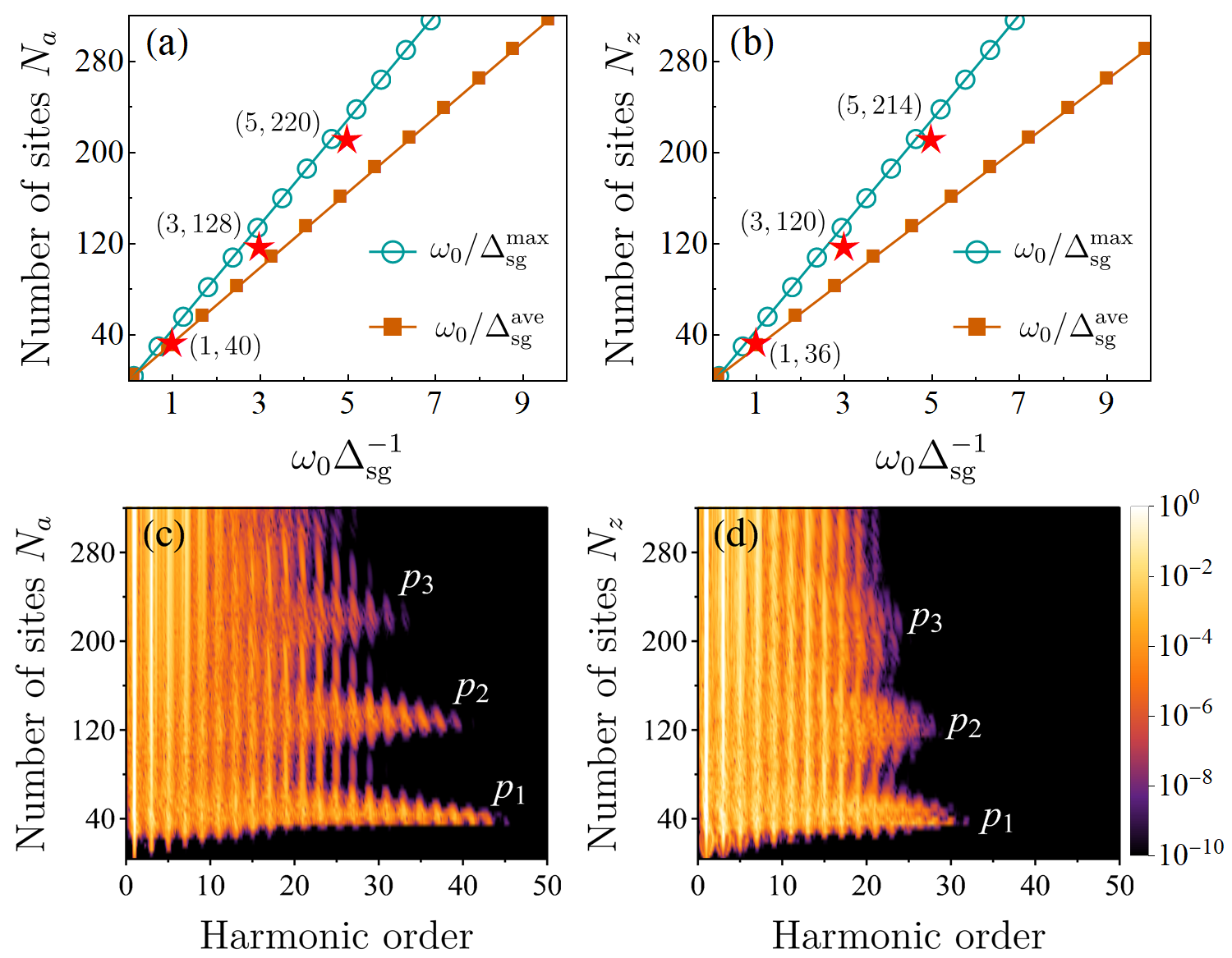}
	\caption{(a)--(b) Ratios of the driver frequency $\omega_{0}$ and subband gaps $\Delta_{\rm sg}$ at Dirac points as a function of system size $N$ for (a) AGNRs and (b) ZGNRs. Lines with symbol $\circ$ ($\blacksquare$) represent the variation of the maximum (average) value of subband gaps. (c)--(d) Harmonic spectra vs number of sites $N$ in (c) AGNRs and (d) ZGNRs, respectively, using a driver with $n_{\text{cyc}}=16$, $\omega_{0}=0.0152\text{ a.u. }(\lambda=3\;\mu\text{m})$, and $E_{0}=0.0012\text{ a.u. }(I_{0}=5.04\times 10^{10}\text{ W/cm}^{2})$. The color bar characterizes the intensity of harmonics. The resonance peaks ($p_{1}, p_{2}$ and $p_{3}$) in (c) and (d) are marked in (a) and (b), respectively, in the form of red stars.}
	\label{sizedepen}
\end{figure}

The subbands near the Dirac points are approximately equidistant. We denote the gap between nearest-neighbor subbands as $\Delta_{\rm sg}$, shown in Figs.~\ref{fig1}(c) and \ref{fig1}(d). It is characterized by $\Delta_{\rm sg}\sim N^{-1}$. We can see in Figs.~\ref{sizedepen}(a) and \ref{sizedepen}(b) that both $(\Delta^{\text{max}}_{\rm sg})^{-1}$ and $(\Delta^{\text{ave}}_{\rm sg})^{-1}$ for AGNR and ZGNR are approximately linear with number of sites $N$, where $\Delta^{\text{max}}_{\rm sg}$ is the maximum subband gap, and $\Delta^{\text{ave}}_{\rm sg}$ is the average value of all subband gaps. Both $\Delta^{\text{max}}_{\rm sg}$ and $\Delta^{\text{ave}}_{\rm sg}$ are presented since the subband gaps at the Dirac points are not precisely equal.

Figures~\ref{sizedepen}(c) and \ref{sizedepen}(d) display the size dependence of harmonic spectra for AGNR and ZGNR, respectively, while the driver field is polarized along the transverse direction. The laser frequency and intensity are fixed at $\omega_{0}=0.41$ eV ($0.0152$ a.u.) and $I_{0}=5.04\times10^{10}$ W/cm$^2$, respectively. In the spectrum of AGNR-HHG as shown in Fig.~\ref{sizedepen}(c), the harmonic cutoff is drastically extended for widths $N_{a}=40$, 128 and 220 (ribbon width $w_{a}\approx 4.8$ nm, 15.6 nm and 26.9 nm), forming three peaks $p_{1}$, $p_{2}$ and $p_{3}$ respectively. These three peaks are depicted in Fig.~\ref{sizedepen}(a) as red stars which abscissas correspond to $\Delta_{\rm sg}=\omega_{0}$, $\omega_{0}/3$ and $\omega_{0}/5$. For the first peak, at $N_{a}=40$, it is found in Fig.~\ref{sizedepen}(a) that $\Delta_{\rm sg}^{\text{ave}}(40)\lesssim\omega_{0}\lesssim\Delta_{\rm sg}^{\text{max}}(40)$, i.e., the laser frequency matches subbband gap [$\omega_{0}\approx\Delta_{\rm sg}(40)$]. Therefore, the harmonic spectrum is affected significantly when the resonance condition ($\omega_{0}=\Delta_{\rm sg}$) is met. Two additional peaks at $N_{a}=128$ and 220 reveal that the resonant excitation occurs not only when $\Delta_{\rm sg}(40)\approx\omega_{0}$, but also when $\Delta_{\rm sg}(N_{a})\approx\omega_{0}/3$ and $\omega_{0}/5$. This can be understood by the optical selection rule of AGNR in Appendix~\ref{sec:selection_rules} when subbands are from the same group. The intraband (interband) transitions are permitted for such two subbands which band index difference $\Delta J=\text{odd}$ ($\Delta J=\text{even}$). Their energy gaps have such a relation: $E_{Js}-E_{J's'}\approx(2j-1)\Delta_{\rm sg}$, where $j$ is an integer. Thus the resonance condition for AGNR can be written as $\Delta_{\rm sg}(N_{a})\approx\omega_{0}/(2j-1)$. Since intraband (interband) transitions are not permitted for $\Delta J =\text{even}$ ($\Delta J =\text{odd}$), we are finally not able to observe such peaks at $N_{a}=84$ and 172 satisfying $\Delta_{\rm sg}(N_{a})\approx\omega_{0}/2j$. Moreover, from $p_{1}$ to $p_{3}$, the peaks gradually widen and their cutoff progressively shrinks, as shown in Fig.~\ref{sizedepen}(c). The former can be attributed to the fact that the subband gap is not very sensitive to the change in system size when the size is large enough. The latter stems from the increase in detuning between the driving frequency $\omega_{0}$ and the excitation gap $(2j-1)\Delta_{\rm sg}$, deviating from the resonance condition. Analogous to AGNR-HHG, the ZGNR-HHG spectrum shows similar features in Fig.~\ref{sizedepen}(d) and can be explained in a same way.

To gain deeper insights into the mechanism of cutoff extension on the harmonic output, we will study the electron dynamics in the presence of the nearly resonant field. We will show how the electrons are excited to the highest conduction subbands emitting high-energy photons.

\section{\label{sec:IV}cascade resonance}
\subsection{\label{sec:IVA}Electron dynamics in GNRs under cascade resonance}

\begin{figure}
	\centering
	\includegraphics[width=\columnwidth]{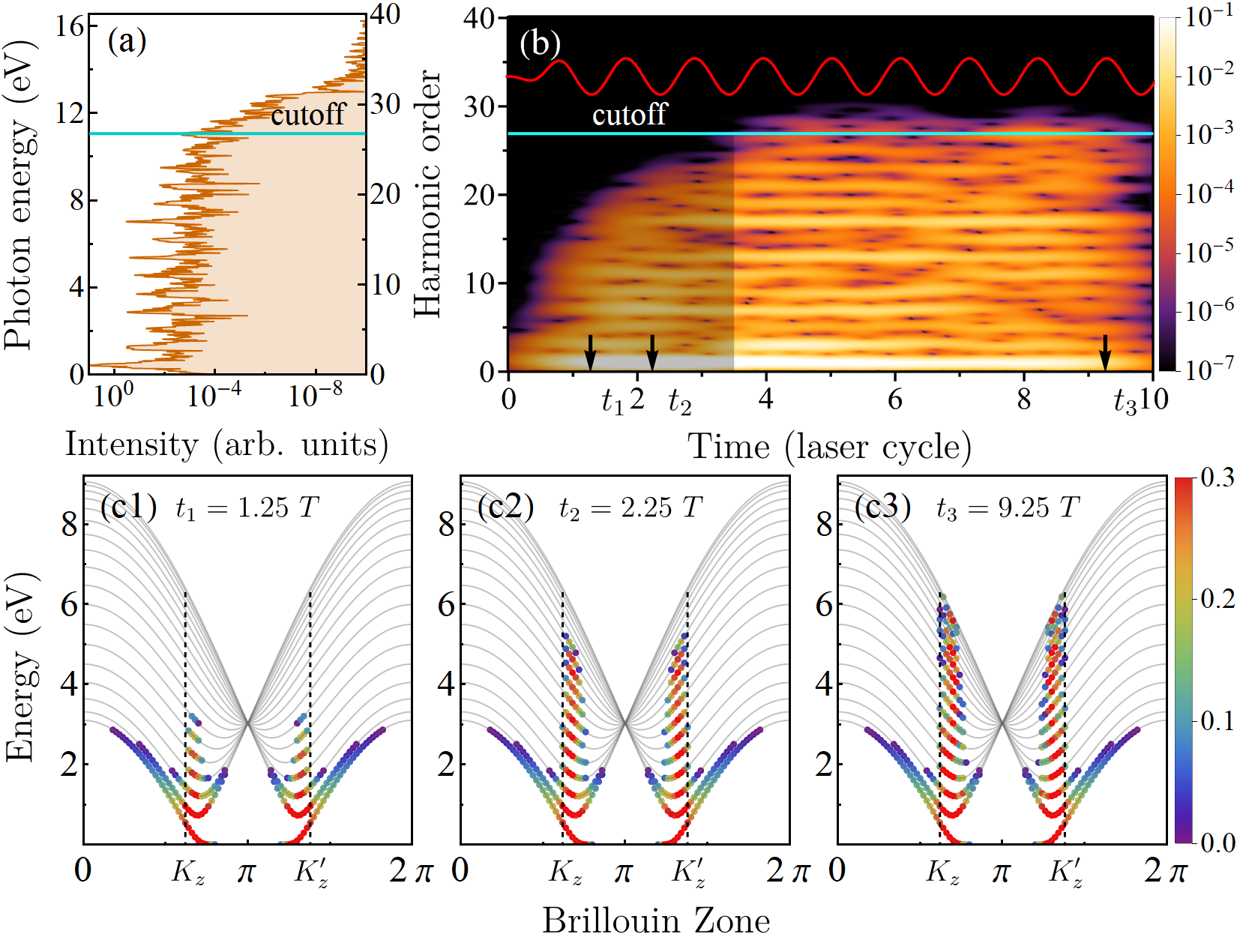}
	\caption{(a) Harmonic spectrum for 36-ZGNR interacting with resonant field. The field strength is $E_{0}=0.0012$ a.u. ($I_{0}=5.04\times 10^{10}$ W/cm$^2$), and the field wavelength is $\lambda=3\;\mu$m ($\omega=0.0152$ a.u.). (b) Time-frequency distribution of the harmonic emission. The red curve is the vector potential of laser field. The cyan line indicates the cutoff frequency. (c1)--(c3) Electron dynamics in the conduction subbands for times $t_i$ marked in (b). $T=2\pi/\omega$ is the optical cycle. The color bar indicates the electron population.}
	\label{zig40edyna}
\end{figure}

Taking $36$-ZGNR as an example, we study the HHG process combined with its electron dynamics under the resonance. To rule out the effect of pulse envelope on quantum paths, a $10$-cycle trapezoid envelope with an one-cycle linear ramp is applied. Figure~\ref{zig40edyna}(a) displays the harmonic spectrum when laser frequency matches the subband gap. It can be seen that the cutoff can reach at the 27th order, whereas such a high-order harmonic cannot be observed if the laser is polarized along the ribbon direction. The time-frequency distribution of the corresponding time-dependent current is presented in Fig.~\ref{zig40edyna}(b), demonstrating the harmonic emission with subcycle temporal resolution. The maximum order of the emitted harmonics increases with the time evolution at the beginning $3.5$ cycles (shadow area) and then keeps at $27$th order for the following $6.5$ cycles. We thus refer to the first $3.5$ cycles as the rise stage and the next $6.5$ cycles as the oscillation stage.

As shown in Figs.~\ref{zig40edyna}(c1)--(c2), after being excited to the conduction band, the electrons move further up to the higher subbands with increasing vector potential $|A(t)|$. When $|A(t)|$ starts to decrease, however, the electrons near the Dirac points seem to do not deexcite to the lower subbands as those electrons governed by Bloch acceleration theorem do. Most of them tend to stay in place and wait for the next half cycle of the laser pulse. Once $|A(t)|$ increases again, these (deposited) electrons will be excited to the higher subbands. As the electrons are excited and then deposited, accumulation zones are formed gradually near the $K_{z}$ and $K'_{z}$. The rise region of the harmonic emission in Fig.~\ref{zig40edyna}(b) is exactly ascribed to the cascade excitation process near the Dirac points. When all the subbands are involved (at about $t=3.5\;T$), see Fig.~\ref{zig40edyna}(c3), the electrons are no longer excited to the higher subbands, but jump up and down in the accumulation zones, thus leading to the oscillation region in Fig.~\ref{zig40edyna}(b). We thus refer to it as cascade resonance. In AGNR, the resonant dynamics is similar. We provide the dynamic images of the conduction band population of $40$-AGNR and $36$-ZGNR in the supplementary material \cite{Suppl}, so as to gain more insight into the electron resonant dynamics. Note that the electron dynamics in valence band is not shown since the distribution of the holes on the valence band is symmetric to the electron distribution on the conduction band.

With a long enough laser pulse, it is not difficult to foresee that the electrons can be driven to the highest subbands, generating high-order harmonics which energy equals the maximum band gap. This explains the significant cutoff extension on the size-dependent harmonic spectra shown in Figs.~\ref{sizedepen}(c) and \ref{sizedepen}(d), when the resonance condition is satisfied.

\subsection{\label{sec:IVB}Analysis of cascade resonance in simplified model}
In order to analyze the cascade resonance near the Dirac points of GNRs and generalize the conclusions to other confined systems, we introduce a $N$-level model here which is widely used to simulate electron dynamics and HHG in quantum dots \cite{Shore1991,Chuquantumdots2010,Wumengxi2016}.
The Hamiltonian reads $H=\sum_{i}^{N}E_{i}\mid i\rangle\langle i\mid$,
where $E_{i}$ is the energy of state $\mid i\rangle$. In the dipole approximation, the time-dependent Hamiltonian of the laser field with the $N$-level quantum dot is written as
\begin{align}
    H(t)=\sum_{i}^{N}E_{i}\mid i\rangle\langle i\mid+ \mathbf{E}(t)\cdot\sum_{i\neq j} \mathbf{d}_{ij}\mid i\rangle\langle j\mid,
    \label{Hqd}
\end{align}
where electric field $\mathbf{E}(t)=E_{0}\cos(\omega t)\hat{y}$, and $\mathbf{d}_{ij}$ is the dipole matrix element between states $\mid i\rangle$ and $\mid j\rangle$. Rabi frequency is defined as $\Omega_{ij}=E_{0}\cdot d^{y}_{ij}$. In principle, the Hamiltonian Eq.~(\ref{Hqd}) can be used to simulate electron dynamics and HHG in any confined system including GNRs along transverse direction, as long as the energy levels and dipole matrix elements are known.

In our simulation, energy levels $E_{1},E_{2},\ldots,E_{N/2}$ are involved as the initial state. Solving the time-dependent Schr{\"o}dinger equation, the dipole moment can thus be evaluated by $\mathbf{d}(t)=\hat{y}\sum_{i,j}\langle\psi(t)\mid i\rangle d^{y}_{ij}\langle j\mid\psi(t) \rangle$. By the Fourier transformation of $\mathbf{d}(t)$, harmonic spectrum is obtained $S(\omega)=\mid\text{FT}[\mathbf{d}(t)]\mid^2$.

\subsubsection{Equi-spaced N-level model}
First of all, we consider an ideal situation: the $N$-level system with equal separations
\begin{align}
E_{i+1}-E_{i}=\Delta_{\rm sg},
\end{align}
where $i=1,2,\ldots,N-1$, and $\Delta_{\rm sg}$ is the gap between energy levels. The dipole matrix elements (in atomic unit) take the form of
\begin{align}
	\begin{cases}
	    d_{ij}=1.0, & \mid i-j\mid=1\\
		d_{ij}=0.3/\mid E_{j}-E_{i}\mid, & \mid i-j\mid\text{is odd but }>1\\
		d_{ij}=0.0, & \mid i-j\mid \text{is even}.
	\end{cases}\label{dipmatelem}
\end{align}
It is constructed based on the features of dipole matrix elements in GNRs, for example, Eq.~(\ref{dmatelem}).

For $N=2$, the model is reduced to the classical Rabi model which has been investigated extensively before. In the resonant case, $\omega=\Delta_{\rm sg}$, the parameter
\begin{align}
	\gamma_{R}=\frac{\Omega_{R}}{\omega}
\end{align}
is commonly adopted to identify different interaction regimes, where $\Omega_{R}=\Omega_{12}=E_{0}\cdot \mid d_{12}\mid$. In the weak coupling regime
$\gamma_{R}\ll 1$, the Rabi flopping can be observed that the population inversion oscillates
between $-1$ and $1$ at frequency $\Omega_{R}$ periodically \cite{Rabi1936}, and thus the
area theorem is valid \cite{McCall1969}. 
While transforming into the strong coupling regime $\gamma_{R}\gtrsim 1$, the area theorem breaks down \cite{Hughes1998,Hughes2000,Ziolkowski1995}. A chaotic oscillation mode displaces the periodic mode since the contribution of counterrotating terms becomes prominent, which is known as the carrier-wave Rabi flopping \cite{Mucke2001,Ciappina2015,Kruchinin2018}.
\begin{figure}
	\centering
	\includegraphics[width=\columnwidth]{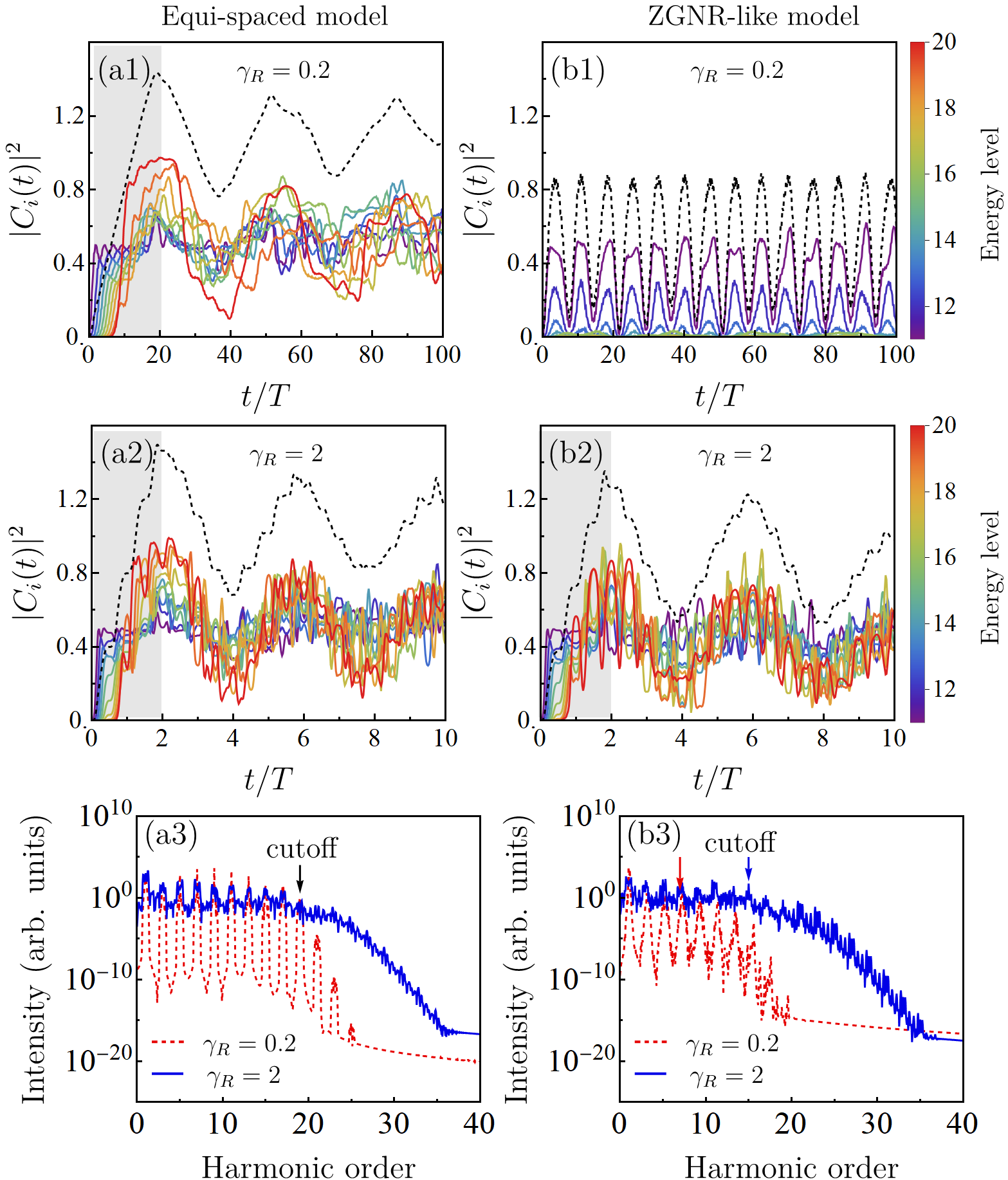}
	\caption{Population dynamics of conduction states in an equi-spaced model with $N=20$ in (a1) the weak coupling regime ($\gamma_{R}=0.2$) and (a2) the strong coupling regime ($\gamma_{R}=2$). Corresponding plots for the $20$-level ZGNR-like model are shown in (b1) and (b2). The color bar identifies the conduction levels and the black dashed curve shows the total population of $i=11, \ldots 20$ scaled by $1/5$ in all cases except (b1).  (a3) and (b3) show the HHG spectrum for the equi-spaced and the ZGNR-like systems, respectively. The red and blue curves represent harmonic spectra for $\gamma_{R}=0.2$ and $\gamma_{R}=2$, respectively.}
	\label{rd}
\end{figure}

In a system with $N>2$, we need to reanalyze its resonant dynamics, because the electron dynamics of the two-level system no longer applies. For comparison, the concepts of weak coupling, strong coupling, and coupling parameter are borrowed. In multilevel systems, we define the coupling parameter as
\begin{align}
    \gamma_{R}=\frac{\Omega_{R}}{\omega}=\frac{\langle \mid d_{ij}\mid\rangle\cdot E_{0}}{\omega},
\end{align}
where $\langle\ldots\rangle$ denotes the average value, $d_{ij}$ is 
the dipole matrix element between the two energy levels which meet the 
resonance condition in frequency, and Rabi frequency $\Omega_{R}=\langle \mid d_{ij}\mid\rangle\cdot E_{0}$. We focus on the case 
$\omega=\Delta_{\rm sg}$ in the following because other resonance cases 
[$\omega=(2j-1)\cdot\Delta_{\rm sg}$, $j>1$] are similar.

Taking $N=20$ as an example, the resonant dynamics on conduction levels is shown in Figs.~\ref{rd}(a1)--(a2). We set the energy separation $\Delta_{\rm sg} =1$ a.u., and the driver frequency $\omega=1$ a.u.. The valence population is not shown because it is symmetric to the conduction population and satisfies $\mid C_{i}\mid^{2}+\mid C_{N+1-i}\mid^{2}=1$, where $\mid C_{i}\mid^{2}$ represents the population on the energy level $E_{i}$. In the weak coupling regime  
$\gamma_{R}=0.2$ ($E_{0}=0.2$ a.u., $\langle d_{i,i+1}\rangle=1.0$ a.u.), see Fig.~\ref{rd}(a1), the excitation process can be divided into two stages: 
rise and oscillation, similar to what has occurred in the 
laser-driving GNR. In the rise region (shadow area), we can find that the electrons are excited to the conduction subbands in a cascaded way. 
Correspondingly, the total conduction population 
($P_{c}=\sum^{N}_{i=N/2+1}\mid C_{i}\mid^{2}$) exhibits a monotonous increase shown by the black dashed line in Fig.~\ref{rd}(a1). Thus the rise time ($T_{r}$) is defined as the period from the beginning of the evolution to the first time that $P_{c}$ reaches the maximum. After entering in the oscillation region, the population of conduction levels starts to collectively oscillate up and down with a roughly uniform period $T_{osc}\approx35\;T$, where $T=2\pi/\omega$. Nevertheless, this periodicity is not strict and the population does not return to the initial value even for long enough pulses. Hence the area theorem fails for cascade resonance even in the weak coupling regime. For strong coupling $\gamma_{R}=2$ ($E_{0}=2$ a.u.), the population dynamics is unaltered compared to the case of $\gamma_{R}=0.2$ a.u., as shown in Figs.~\ref{rd}(a2). The only significant change is that the oscillation period $T_{osc}$ becomes $3.5\;T$. Therefore, the regime transformation in the equi-spaced system does not fundamentally shift the behavior of resonant excitation. This is completely different from the two-level resonance.

Based on the above analysis, it is found that the cascade resonance can take place in equi-spaced $N$-level systems as long as the frequency conditions are met, no matter how weak the electric field is. Due to all the energy levels are involved in the cascade resonance, the cutoffs of harmonic spectra can thus extend to maximum energy gap $(E_{20}-E_{1})$. Therefore, we can see from Fig.~\ref{rd}(a3) that the harmonic cutoffs reach $19$th order [$(E_{20}-E_{1})/\omega=19$] for both $\gamma_{R}=0.2$ and $2$, and the intensities of corresponding harmonic spectra are comparable.

Moreover, the rise time $T_{r}$ is a continuous function of the Rabi frequency $\Omega_{R}$ and the number of energy levels $N$. In Appendix~\ref{appenrise}, Figures~\ref{risetime}(a) and \ref{risetime}(b) clearly show that the rise time exhibits a linear dependence of $\Omega^{-1}_{R}$ and $N$. Mathematically, we describe it by
\begin{align}\label{tr}
	T_{r}\approx \delta\cdot N/\Omega_{R}=\delta\cdot N/(\langle\mid d_{ij}\mid\rangle\cdot E_{0}),
\end{align}
where $\delta$ is the scale factor. This equation indicates a fact that the weaker the field strength and the more the energy levels, the longer the rise time. Reflected in the HHG process of the GNR, it implies that the larger the size and the weaker the field strength, the longer the rise region in harmonic emission.

\subsubsection{ZGNR-like N-level model}

Now we turn to a more realistic situation taking into account the detuning between the driver frequency and the energy separation. Simulating the subband structure of ZGNR at the Dirac points, energy levels have the form:
\begin{align}
	E_{i+1}-E_{i}=\Delta_{\rm sg}\left[f-g\left(\frac{i-N/2}{N/2}\right)^{2}\right],
\end{align}
where $f$ and $g$ are the parameters which determine the nearest-neighbor energy difference. In the following simulation, we set $N=20$, $\omega=\Delta_{sg}=1$ a.u., $f=1.15$, and $g=1.1$. The energy difference near the Fermi level is $E_{11}-E_{10}=1.15\;\omega$, slightly deviating from the driver frequency. The dipole matrix elements we used the same as in Eq.~(\ref{dipmatelem}).

In this ZGNR-like system, we focus on the population dynamics in different coupling regimes as before. In the weak coupling regime ($\gamma_{R}=0.2$), see Figs.~\ref{rd}(b1), the cascade resonance disappears but is displaced by a behavior close to the classical Rabi
flopping. This is called near resonance. Only four of conduction levels near the Fermi level are
involved, which implies that only a small proportion of the electrons can be driven to the highest energy level. 
Therefore, the corresponding HHG spectrum (red dashed curve) in Fig.~\ref{rd}(b3) terminates at $7$th order which energy is much lower than the maximum gap [$(E_{20}-E_{1})/\omega= 15.37$].
However, in the strong coupling regime ($\gamma_{R}=2$), the collective
resonance reappears in the ZGNR-like system, as shown in Figs.~\ref{rd}(b2). The rise and oscillation stages constitute the entire excitation process again. Correspondingly, the cutoff of harmonic spectrum can thus extend to $15$th order shown as the blue curve in Fig.~\ref{rd}(b3). Therefore, the cascade resonance takes place merely for $\gamma_{R}\gtrsim 1$ when the field
detuning exists. This is different from the population dynamics in the
equi-spaced model.

\subsection{Resonance conditions for laser-driving GNRs}

Combining the analyses of resonant dynamics in the 36-ZGNR (Sec.~\ref{sec:IVA}) and in the
simplified models (Sec.~\ref{sec:IVB}) together, it is natural to infer that one of the condition for the occurrence of cascade resonance in the laser-driving GNRs
is $\gamma_{R}\gtrsim1$. We verify the coupling parameter at $p_{1}$ in the ZGNR-HHG spectrum
[Fig.~\ref{sizedepen}(d)], and find that $\gamma_{R}(N_{z}=36)\approx1.3>1$ at the Dirac points.

Thereby, the resonance conditions for
a laser-driving GNR can be summarized as
\begin{subequations}
	\begin{align}
		\label{rescda}
		\Delta_{\rm sg}(N)&\approx\omega/(2j-1),\\
		\label{rescdb}
		\gamma_{R}(N)&=E_{0}\cdot\langle \mid d_{\Delta J}^{y}(N)\mid\rangle/\omega\gtrsim1,
	\end{align}
\end{subequations}
where $j$ is an integer, and $\Delta J$ is the index difference of resonant subbands. Equation~(\ref{rescda}) is the frequency condition we have obtained in Sec.~\ref{sec:III}. It determines which driver frequency can lead to cascade resonance in a GNR with size of $N$. Then equation~(\ref{rescdb}) states that the occurrence of cascade resonance requires the laser-driving GNR to be in the strong coupling regime, and $\gamma_{R}=1$ is the critical point. Further, we define a concept --- critical field strength:
\begin{align}\label{crit}
	E_{\rm cri}(\Delta J,N)=\omega/\langle \mid d_{\Delta J}^{y}(N)\mid\rangle,
\end{align}
which is the minimum field strength resulting in the cascade resonance in a $N$-GNR. Substituting Eq.~(\ref{rescda}) into Eq.~(\ref{crit}), the critical field strength can be evaluated by
\begin{align}\label{crital}
	E_{\text{cri}}(\Delta J,N)=(2j-1)\cdot\Delta_{\rm sg}(N)/\langle \mid d_{\Delta J}^{y}(N)\mid\rangle.
\end{align}

When $j=1$, $\Delta_{\rm sg}$ including $\Delta_{\rm sg}^{\rm max}$ and $\Delta_{\rm sg}^{\rm ave}$ is a linear function of $N^{-1}$ [Figs.~\ref{sizedepen}(a) and \ref{sizedepen}(b)], and $\langle\mid d_{\Delta J=1}^{y}(N)\mid\rangle$ is linear with $N$ (Fig.~\ref{fig2}). Substituting the linear fitting results of $\Delta_{\rm sg}^{\rm max}$ and $\langle\mid d_{\Delta J=1}^{y}(N)\mid\rangle$ into Eq.~(\ref{crital}), we finally obtain $E_{\text{cri}}(N)$ as a function of $N$,
\begin{align}
	\begin{cases}
		E_{\text{cri}}^{a}(N_{a})=(\alpha_{a} N_{a}^{2}+\beta_{a} N_{a}+\gamma_{a})^{-1}, & \text{AGNR}\\
		E_{\text{cri}}^{z}(N_{z})=(\alpha_{z} N_{z}^{2}+\beta_{z} N_{z}+\gamma_{z})^{-1}, & \text{ZGNR},
	\end{cases}
\end{align}
where $(\alpha_{a},\beta_{a},\gamma_{a})\approx(0.025,0.039,0.014)$, and $(\alpha_{z},\beta_{z},\gamma_{z})\approx(0.021,0.04,0.014)$. It exhibits a dependence of $E_{\text{cri}}(N)\sim N^{-2}$ in the large $N$ limit. It clearly indicates that the cascade resonance induced harmonic cutoff extension is more easily observed in large size GNRs.

\subsection{\label{sec:gappedZGNR}Electron dynamics and HHG away from resonance condition in GNRs}

\begin{figure}[b]
	\centering
	\includegraphics[width=\columnwidth]{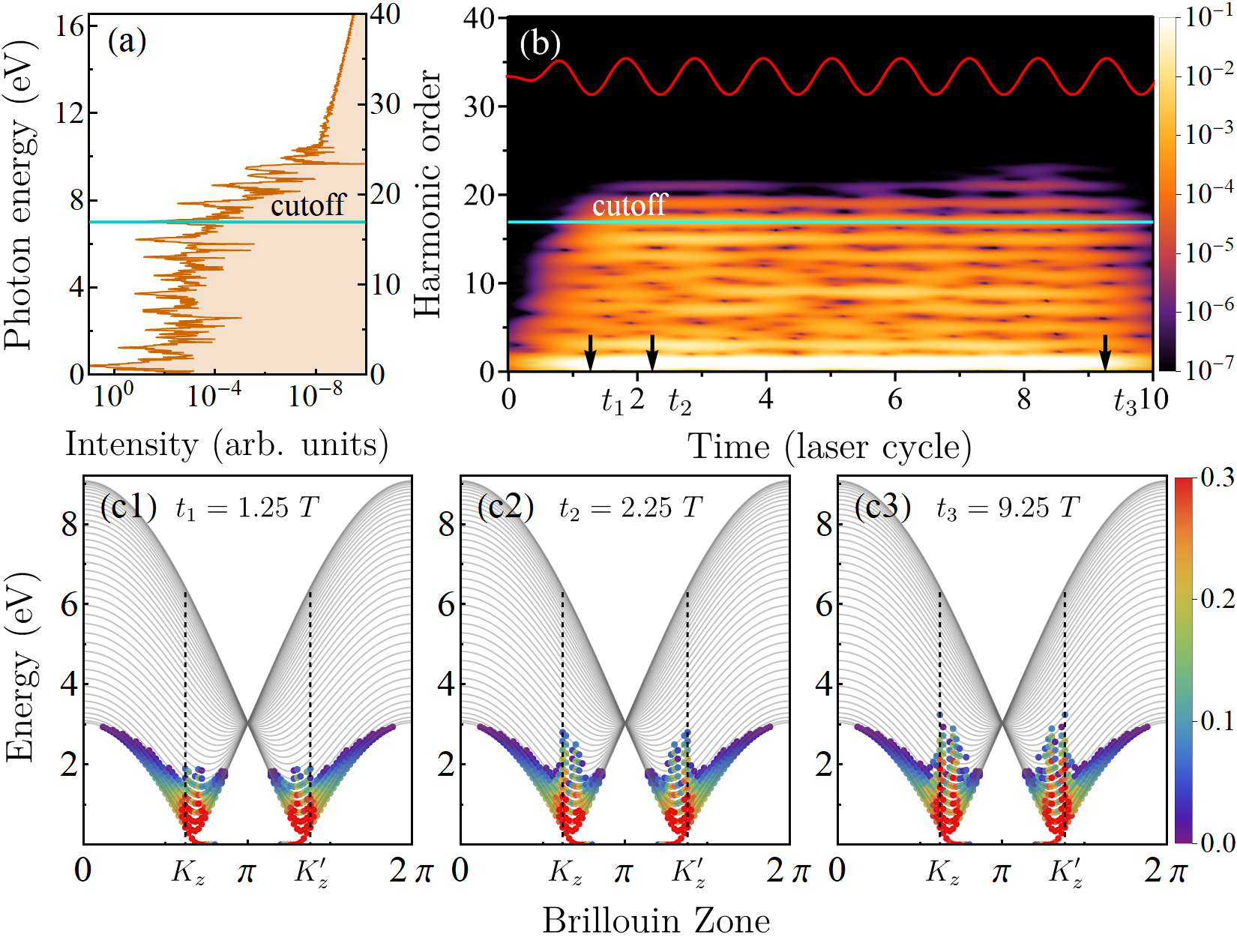}
	\caption{(a) Harmonic spectrum for 80-ZGNR. The field strength is $E_{0}=0.0012$ a.u. ($I_{0}=5.04\times 10^{10}$ W/cm$^2$), and the field wavelength is $\lambda=3\;\mu$m ($\omega=0.0152$ a.u.). (b) Time frequency distribution of the harmonic emission. The red curve is the vector potential of external field. The cyan line indicates the cutoff frequency. (c1)--(c3) Electron dynamics in the conduction subbands for the times $t_{i}$ marked in (b).}
	\label{eledmismatch}
\end{figure}
A natural question at this stage is how sensitive is the cascade resonance to the frequency-matching and field-strength conditions.

In order to examine the frequency-matching condition, we choose the $80$-ZGNR ($\Delta_{\rm sg}\approx 0.2$ eV) but use the laser field with wavelength of $3$ $\mu$m ($\omega\approx 0.41$ eV) which causes cascade resonance in $36$-ZGNR ($N_{z}=36$) to study the HHG process.
In Figs.~\ref{eledmismatch}(a) and \ref{eledmismatch}(b), the plateau of harmonic spectrum merely extends to 17th
order, which is evidently shorter than that in $36$-ZGNR (27th order). In Figs.~\ref{eledmismatch}(c1)--(c3), it is found that the electron dynamics is far from the cascade resonance. On the subbands with energies above $3.5$ eV, the population is not as
significant as in the case of a cascade resonance. This is because the relation between subband gap and laser frequency ($\Delta_{\rm sg}\approx\omega/2$) does not agree with the frequency condition in Eq.~(\ref{rescda}). Nevertheless, the subband gaps are not uniform. The gaps near the Fermi energy are larger than those higher, which can satisfy the frequency condition to some extent. Therefore, we observe that subbands close to the Fermi energy are involved in a near resonance. 
The cascade resonance for $80$-ZGNR, however, appears if we employ a laser with wavelength $\lambda=6\;\mu$m ($\omega\approx0.21$ eV) satisfying the frequency condition ($\Delta_{\rm sg}\approx\omega$). The accumulation zones of electrons span from $0$ to $6.5$ eV and harmonic cutoff exceeds $11$ eV  (not shown).

When the laser field does not reach the threshold strength [Eq.~(\ref{crital})], the electron dynamics in GNRs is similar to what has been explored in the ZGNR-like model, see Sec.~\ref{sec:IVB}. The electrons are not able to be excited to high-energy subbands like what has been presented in Figs.~\ref{eledmismatch}(c1)--(c3). The intensity of harmonic spectrum is lower and the plateau is shorter than those in the strong laser field. In next section, the study of cutoff law on the field
strength systematically shows the variation of harmonic
spectrum from weak coupling regime to strong coupling regime.

\section{\label{sec:V}Cutoff law of HHG in graphene nanoribbons}

The cutoff law of HHG is a fundamental issue in strong-field physics. In GNRs, it has been shown that the energy cutoff increases linearly with increasing field strength and wavelength for fields along the ribbon direction, as also shown in the Appendix \ref{ribbonHHG}. This is consistent with the generic observations in bulk solids. When the laser polarization turns to the cross-ribbon direction, however, the accelerated motion of electrons on the energy bands switches to the electron dynamics for finite systems like cascade resonance, and thus leading to alteration of the cutoff law as well. In the following, we explore the cutoff law of GNR-HHG on the laser duration, field strength and wavelength.
\begin{figure}
	\centering
	\includegraphics[width=\columnwidth]{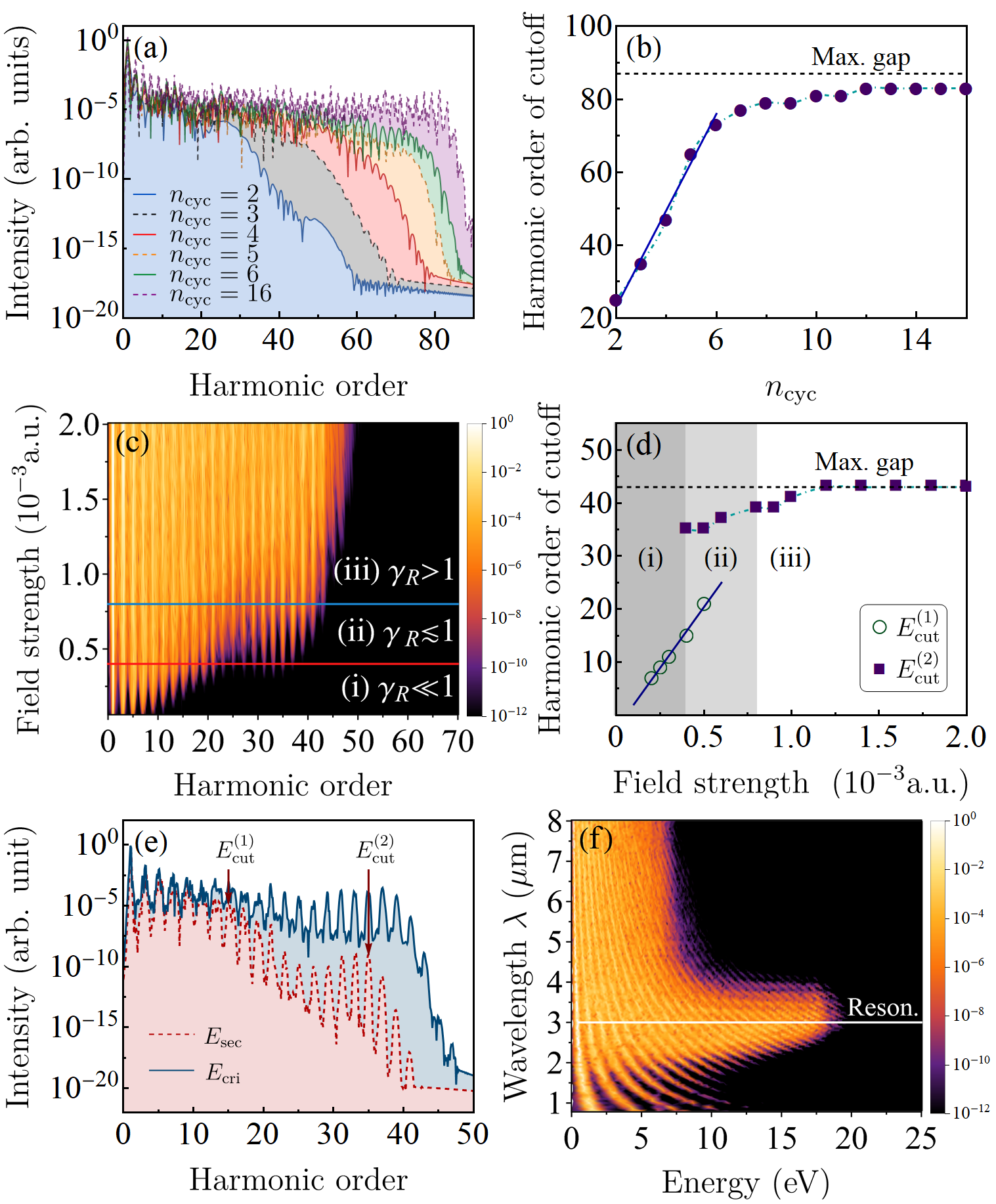}
	\caption{(a) Harmonic spectra vs number of cycles ($n_{\text{cyc}}$) in $80$-AGNR. The field strength is $E_{0}=5\times 10^{-4}$ a.u. ($I_{0}=8.75\times 10^9$ W/cm$^2$), $\gamma_{R}\approx1.25$, and the frequency is $\omega_{0}=0.0076\text{ a.u. }(\lambda=6\;\mu\text{m})$. (b) The cutoff as a function of $n_{\rm cyc}$ corresponding to the harmonic spectra in (a). (c) Harmonic spectra vs field strength in $40$-AGNR, using a driver with $n_{\text{cyc}}=16$, and $\omega_{0}=0.0152\text{ a.u. }(\lambda=3\;\mu\text{m})$. The color bar indicates the intensity of harmonics. (d) The cutoff as a function of field strength, corresponding to the harmonic spectra in (c). (e) Harmonic spectra extracted from (c) for $E_{\text{sec}}=4\times10^{-4}$ a.u. and $E_{\text{cri}}=8\times10^{-4}$ a.u.. (f) Harmonic spectra vs wavelength in a $40$-AGNR, using a driver with $n_{\text{cyc}}=16$, and $E_{0}=0.0012\text{ a.u. }(I_{0}=5.04\times 10^{10}\text{ W/cm}^{2})$. The white line represents the onset of cascade resonance.}
	\label{cutoff}
\end{figure}

{\em Duration. ---} Figure~\ref{cutoff}(a) shows the HHG spectra in the $80$-AGNR subjected to resonant laser pulses with different durations. In the calculation, laser pulses in form of Eq.~(\ref{laser}) are employed, so the number of laser cycles ($n_{\text{cyc}}$) can be used to indicate pulse duration. We note that the plateau of harmonic spectrum is gradually stretched with increasing the laser duration. In Fig.~\ref{cutoff}(b), the harmonic cutoff clearly exhibits linear scaling with pulse duration for $n_{\text{cyc}}\leqslant5$ and reaches the maximum attainable energy at $n_{\text{cyc}}=11$. Between $n_{\text{cyc}}=6$ and $n_{\text{cyc}}=10$, the cutoff varies slowly with increasing the laser duration. This cutoff law can be interpreted qualitatively by the two-stage excitation process in GNRs under the cascade resonance as elaborated in Secs.~\ref{sec:IVA} and \ref{sec:IVB}. The linear increase and the slow variation of the cutoff corresponds to the rise and oscillation stages in the resonant excitation, respectively.

{\em Field strength. ---} In Fig.~\ref{cutoff}(c), we study the field strength dependence of harmonic emissions in the $40$-AGNR interacting with resonant pulses. The blue line represents the critical field strength for the $40$-AGNR in which $E_{\text{cri}}=8\times10^{-4}$ a.u. (i.e., $I_{\text{cri}}=2.24\times10^{10}$ W/cm$^2$), and the red line is located at the field strength ($E_{\text{sec}}=4\times10^{-4}$ a.u.) in which harmonic spectrum starts to possess marked second plateau. Given these two lines, the spectrum is intuitively divided into three regions: (i) $\gamma_{R}\ll1$, (ii) $\gamma_{R}\lesssim1$, and (iii) $\gamma_{R}>1$. (i) In the weak coupling regime ($\gamma_{R}\ll1$), the energy cutoff presents a linear dependence on the field strength, see Figs.~\ref{cutoff}(c) and \ref{cutoff}(d). (ii) When entering the transition regime ($\gamma_{R}\lesssim1$), we can observe two-plateau structures on HHG spectra like the red dashed curve in Fig.~\ref{cutoff}(e), which result from the difference of population on lower and higher subbands. This difference can be understood by the similar electron dynamics in the ZGNR-like system, that is, the lower subbands participate in collective resonance, whereas the higher subbands are in near resonance or even in non-resonance. As Fig.~\ref{cutoff}(d) shows, the cutoff of the first plateau ($E^{(1)}_{\text{cut}}$) preserves the linear dependence in region (i), but the second cutoff ($E^{(2)}_{\text{cut}}$) approaches the attainable maximum tardily. (iii) As the field strength increases up to $E_{\text{cri}}$, the blue curve in Fig.~\ref{cutoff}(e) shows that the first and the second plateaus merge thogether since all subbands are involved in the collective resonance. For this reason, the energy cutoff in the strong coupling regime ($\gamma_{R}>1$) tends to saturate and remains at the maximum band gap. We have thus revealed that the cutoff law of HHG in GNRs depends strongly on the interaction regime to which the system is subjected, the cutoff neither following the linear scaling in electric field of the bulk solids nor the quadratic dependence in electric field of gas.

{\em Wavelength. ---} Figure~\ref{cutoff}(f) shows the HHG spectrum as a function of driver wavelength in a $40$-AGNR ($\Delta_{\rm sg}\approx0.4$ eV). It can be found that the cutoff frequency is extended abruptly around wavelength $\lambda=3\;\mu\text{m}$ ($\omega\approx0.41$ eV) where the frequency condition is satisfied exactly. It is because the cascade resonance appears as the wavelength approaches $3\;\mu\text{m}$. When the driver wavelength increases away from $3\;\mu\text{m}$, the laser-driving AGNR begins to deviate from the resonance point, and thus the plateau of the harmonic spectrum shrinks drastically.

\section{Discussion of cascade resonance in general confined systems}

\begin{figure*}
	\centering
	\includegraphics[width=1.8\columnwidth]{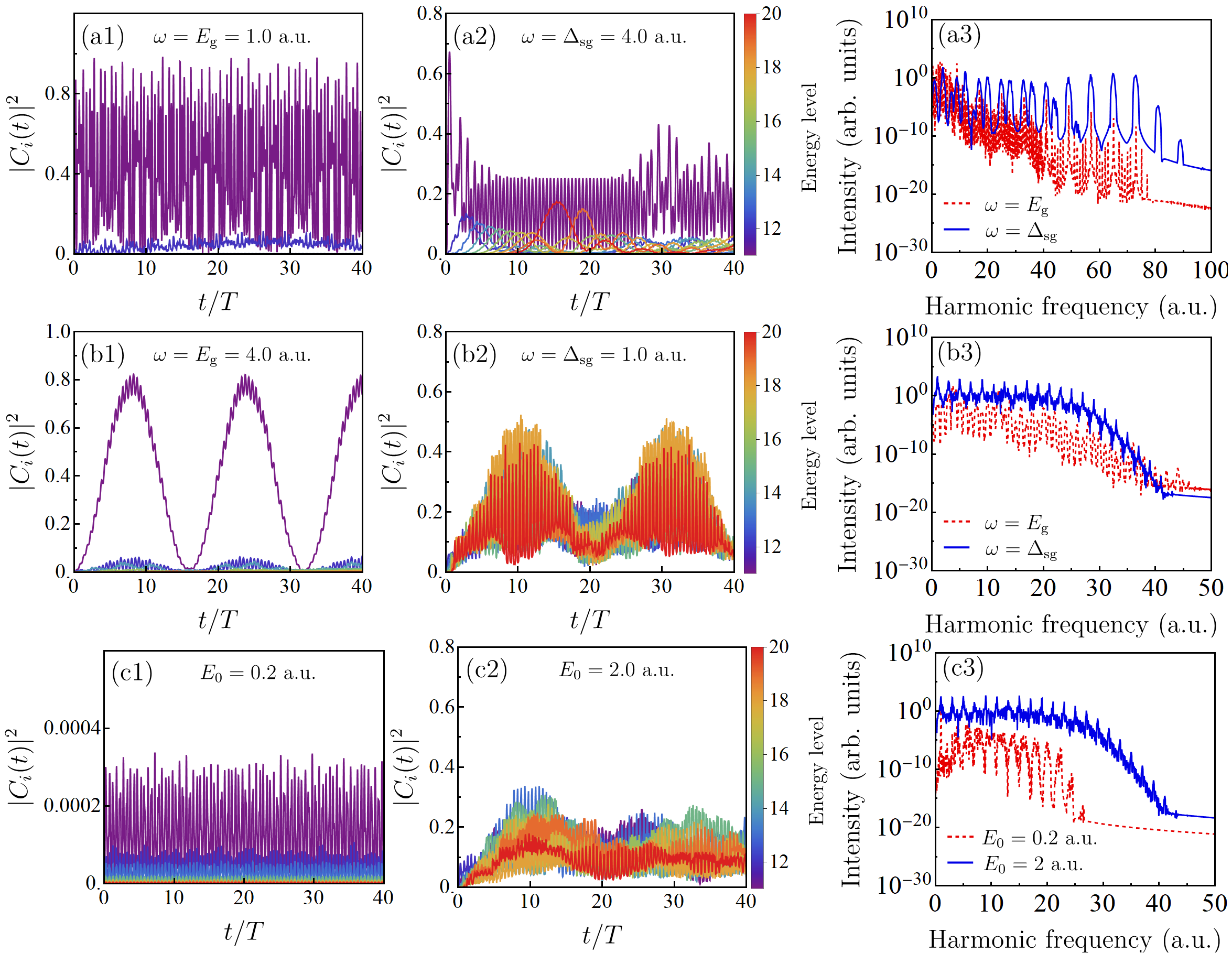}
	\caption{Population dynamics and HHG spectrum depending on energy-level structure. (a1)--(a2) Population dynamics for $E_{\rm g} < \Delta_{\rm sg}$. (a1) laser frequency matches fundamental gap ($\omega =E_{\rm g}= 1$ a.u.). (a2) laser frequency matches sub(band) gap ($\omega =\Delta_{\rm sg}= 4$ a.u.). (a3) Harmonic spectra in red and blue correspond to the electron dynamics in (a1) and (a2), respectively. (b1)--(b2) Population dynamics for $E_{\rm g} > \Delta_{\rm sg}$. (b1) laser frequency matches fundamental gap ($\omega =E_{\rm g}= 4$ a.u.). (b2) laser frequency matches sub(band) gap ($\omega =\Delta_{\rm sg}= 1$ a.u.). (b3) Harmonic spectra in red and blue correspond to the electron dynamics in (b1) and (b2), respectively. (c1)--(c2) Population dynamics for a system with random sub(band) gap in (c1) weak coupling regime ($E_{0}=0.2$ a.u.) and (c2) strong coupling regime ($E_{0}=2$ a.u.). The laser frequency matches average sub(band) gap. (c3) Harmonic spectra in red and blue correspond to the electron dynamics in (c1) and (c2), respectively.
    }
	\label{sens}
\end{figure*}

We now turn our attention to the possibility of observing the cascade resonance induced HHG spectra in general confined
systems. The key ingredients for cascade resonance, \textit{viz.}, the frequency-matching and field strength conditions, can be approximately met in a variety of systems in a nanoribbon geometry \cite{PhysRevB.73.165319}, for example in the Kagmoe lattice \cite{PhysRevB.102.045151} and $\alpha-T_3$ lattice \cite{Zuber2021,Alam_2019} systems. Additionally, quantum dots offer a high degree of tunability in terms of the the energy spectra via the synthesis route, size dependence and/or the external parameters. In general, however, for most materials/systems not only the frequency-matching condition over a large number of subbands
may not be possible, but also the fundamental gap ($E_{\rm g}$) could be significantly different from the
(average) subband gap ($\Delta_{\rm sg}$). 

To explore such systems for possible cascade resonance induced HHG phenomena, we resort to the simplified models akin to the ones
introduced in Sec. \ref{sec:IVB}, but the dipole matrix element takes the form of 
\begin{align}
	\begin{cases}
		d_{ij}=1.0/\mid E_{j}-E_{i}\mid, & \mid i-j\mid\text{is odd}\\
		d_{ij}=0.0, & \mid i-j\mid \text{is even}.
	\end{cases}\label{dme}
\end{align}
We consider $E_{\rm g} \ne \Delta_{\rm sg}$ and laser frequency matched to either the fundamental gap or the average sub(band) gap. 
Therefore, we need two parameters
\begin{align}
    &\gamma_{E}=d_{E_{\rm g}}\cdot E_{0}/\omega=E_{0}/(\omega\cdot E_{{\rm g}})\,\\
    &\gamma_{\Delta}=\langle d_{i,i+1}\rangle\cdot E_{0}/\omega=E_{0}/(\omega\cdot\langle\Delta_{{\rm sg}}\rangle)\,
\end{align}
denoting the coupling strength for fundamental gap and sub(band) gap, respectively.
We additionally consider cases where $\Delta_{\rm sg}$ is non-uniform.
To illustrate these cases, we also consider a $20$-level model.

\textit{Case I: $E_{\rm g} < \Delta_{\rm sg}$ --- }  For brevity, we consider the fundamental gap $E_{\rm g}=1$ a.u. 
while all other energy levels are equally-spaced in energy with a larger gap:
\begin{subequations}
    \begin{align}
       & E_{i+1} - E_{i} = E_{\rm g} =1\; \text{a.u.}\;\quad \,\, \, \text{for } i=N/2,\\
       & E_{i+1} - E_{i} = \Delta_{\rm sg} = 4\; \text{a.u.}\;\quad \,\, \,  \text{otherwise}.
    \end{align}
\end{subequations}
When the laser frequency matches with the fundamental gap $E_{\rm g}$, the system exhibits near resonance irrespective of the coupling strength. The electron dynamics in the strong coupling ($E_{0}=2 \text{ a.u., }\gamma_{E}=2$) is shown in Fig.~\ref{sens}(a1). Significant electron population and oscillation merely occur at the energy levels close to the Fermi energy.
On the other hand, when the laser frequency matches the sub(band) gap $\Delta_{\rm sg}$, the cascade resonance can be achieved, in principle, even for the weak coupling between the sub(band) levels, as shown in Fig. \ref{sens}(a2) for $\gamma_{\Delta}=0.125$ ($E_{0}=2$ a.u.). The population on each energy level is comparable.

The corresponding HHG spectra for different tuning of the laser frequency are shown in Fig. \ref{sens}(a3) We clearly see that the HHG spectrum for the cascade resonance when $\omega = \Delta_{\rm sg}$ is much intense than for the near resonance when $\omega = E_{\rm g}$. The frequency cutoff of resonant harmonic spectrum reaches 72 a.u. which is close to maximum band gap (73 a.u.).

\textit{Case II: $E_{\rm g} > \Delta_{\rm sg}$ ---} We set the fundamental gap and sub(band) gap as
\begin{subequations}
    \begin{align}
       & E_{i+1} - E_{i} = E_{\rm g} =4\; \text{a.u.}\;\quad \,\, \, \text{for } i=N/2,\\
       & E_{i+1} - E_{i} = \Delta_{\rm sg} = 1\; \text{a.u.}\;\quad \,\, \,  \text{otherwise}.
    \end{align}
\end{subequations}
Figure~\ref{sens}(b1) shows the population dynamics for the laser frequency matching the
fundamental gap $E_{\rm g}$. The field strength is set by $E_{0}=1.0$ a.u. ($\gamma_{E}=0.625$). The near resonance which has been observed in Fig.~\ref{sens}(a1) occurs in the energy levels near the Fermi level. Therefore, in the corresponding HHG spectrum, see Fig.~\ref{sens}(b3), the intensity is relatively lower and the plateau structure is not clear. On the other hand, when the laser frequency matches the sub(band) gap, a cascade resonance can occur only when $\gamma_{\Delta}\gtrsim 1$. Figure~\ref{sens}(b2) shows the electron dynamics when $E_{0}=2$ a.u. ($\gamma_{\Delta}=2$). The cascade resonance takes place. The harmonic cutoff, blue curve in Fig.~\ref{sens}(b3), can thus reach maximum energy gap at $22$ a.u. ($E_{20}-E_{1}$).

We further note that the
cascade resonance depends crucially on the ratio of $E_{\rm g}/\Delta_{\rm sg}$.
For much larger fundamental gap values, the electron density in the higher subbands decreases as the excitation of electrons to the first conduction level is probabilistically low. The cascade resonance then cannot take place. In the specific context of GNRs, this can be achieved by adding staggered onsite potentials which open or enlarge the fundamental gap. In Appendix~\ref{sec:gappedZGNR}, we show the HHG spectrum and population dynamics in $36$-ZGNR with a $1.0$ eV gap. We can still observe the cascade resonance near the Dirac points but the lower conduction population in comparison to that of gapless $36$-ZGNR. The intensity of corresponding HHG spectrum is lower than gapless ZGNR-HHG spectrum. However, the cascade resonance disappears when the fundamental gap is larger than 1.5 eV (not show).

A more realistic scenario in materials is when the sub(band) gaps are non-uniform. To account for these
effects, we consider a model where $\Delta_{\rm sg}$ is chosen randomly (uniform probability
distribution) in a given energy range, while $E_{\rm g} > \langle \Delta_{\rm sg}\rangle$, where $\langle
\ldots \rangle$ denotes the average value. The energy levels are: 
\begin{subequations}
    \begin{align}
        & E_{i+1} - E_{i} = E_{g} = 4 \text{ a.u.}\,,\;\qquad \qquad \quad  \text{if}\; i=N/2\\
       & E_{i+1} - E_{i} = \Delta_{\rm sg}= \Delta\cdot \text{Ran}(0.5,1.5)\,,\;\, \text{otherwise}
    \end{align}
\end{subequations}
where Ran$(n,m)$ means a random number between $n$ and $m$. The laser frequency matches the average sub(band) gap ($\omega=\Delta=1$ a.u.).

Figure~\ref{sens}(c1) shows the electron dynamics in the weak coupling regime, where $E_{0}=0.2$ a.u. ($\gamma_{\Delta} \approx 0.2$). Because of the large fundamental gap ($E_{\rm g} = 4$ a.u.), the electrons can hardly be excited to conduction levels. The population of the first conduction level is even less than $0.01$.
In comparison, in the strong coupling regime ($E_{0}=2$ a.u.), shown in Figure~\ref{sens}(c2), all the higher lying energy levels have sizable population. The details of the
energy-level distribution seem immaterial to the cascade resonance phenomena. These features are also clearly reflected in the HHG spectra. In Fig.~\ref{sens}(c3), the intensity of harmonic spectrum in strong coupling regime is much higher than that in weak couling regime.

Consequently, the cascade resonance can be anticipated in real nanoribbon materials where a few subbands are nearly equi-spaced in energy while other bands are randomly distributed. When laser frequency matches subbband gap, the resonant excitation allows more electrons to occupy higher conduction subbands, which eventually enhances and broadens the harmonic spectra in experimental measurements.

\section{Conclusion \& Outlook}
In conclusion, we demonstrated theoretically that the cascade resonance provides a systematic
way to extend and enhance the harmonic spectrum of a confined quantum system. The cascade resonance can pump electrons to higher subbands, resulting in
enhanced higher-order harmonic emissions, which is, therefore, 
fundamentally distinct from the high-harmonic generation mechanism of atoms, molecules and bulk. 
    
Based on the study of size-dependent GNR-HHG and dynamic analysis in a multilevel model, the
resonance conditions for laser parameters like frequency and field strength are established. The cascade resonance occurs when the laser 
frequency matches the subband gap and the field exceeds the threshold strength. 
While large deviations from the ideal frequency-matching condition leads to
disappearance of the cascade resonance, for small deviations, strong field strengths can still induce cascade resonance although the strength of the HHG spectra becomes relatively weak. 

These predictions are well-within the experimental reach, and relatively straight forward to verify in present experimental setups. These ideas can also be applicable to other confined materials/systems which, in general, may not meet the frequency matching condition perfectly. Perhaps, the most interesting situation is when only a part of the subbands meet the frequency matching condition, where the cascade resonance of few levels may take place by carefully tuning the laser frequency and intensity, for example in other two-dimensional materials in a nanoribbon geometry withstanding, or in quantum dots. Among these, the nanoribbons based on kagome lattice and $\alpha-T_3$ lattice systems are particularly appealing due to the presence of flat band together with the linearly dispersing Dirac bands. A detailed study of HHG phenomena in such system is, however, beyond the scope of this work.

Looking forward, our theoretical framework for attosecond physics of confined systems establishes cascade resonance as a powerful tool for obtaining intense simple ultraviolet/extreme-ultraviolet X-ray sources. Attosecond technology has previously focused on the modulation of ultrafast processes by the intensity, wavelength, polarization, and time delay (two pulses) of the incident laser. The resonance mechanism will provide a more diverse manipulation approach because of the sensitivity to material size and drive duration, which may finally allow the establishment of highly tunable solid-state XUV sources. Furthermore, the relation between HHG and cascade resonance in the nanoribbon system will provide a new platform and idea for the study of carrier-wave Rabi flopping.

\section*{Acknowledgements}
Xiao Zhang acknowledges fruitful discussions with Jinbin Li. The authors acknowledge supported from the National Natural Science Foundation of China (NSFC) (Grants No. 11834005, No. 11874030, No.11904146, No. 12047501).

\appendix

\section{\label{sec:methods}Methods}
The atomic structures of GNRs with armchair edges and zigzag edges are presented in
Fig.~\ref{fig1}(a) and \ref{fig1}(b), respectively. The distance of unit cell for AGNR and ZGNR are
$d_{a}=\sqrt{3}a$ and $d_{z}=a$, respectively, where the lattice constant $a=2.46\text{ \AA}$.

\subsection{\label{sec:tb}Tight-binding model for graphene nanoribbon}
The tight-binding Hamiltonian for AGNR takes the form
\begin{equation}
\begin{split}
\hat{H}^{a}=&-\gamma\sum_{i=1}^{N_{x}}\sum_{j=1}^{N_{a}/2}\left(\hat{p}_{i,2j}^{\dagger}\hat{p}_{i,2j-1}+\hat{p}_{i,2j}^{\dagger}\hat{p}_{i,2j+1}\right.\\
&+\hat{q}_{i,2j-1}^{\dagger}\hat{q}_{i,2j}+\hat{q}_{i,2j-1}^{\dagger}\hat{q}_{i,2j-2}+\hat{p}_{i,2j}^{\dagger}\hat{q}_{i,2j}\\
&\left.+\hat{q}_{i,2j-1}^{\dagger}\hat{p}_{i+1,2j-1}\right)+\text{H.c.},
\end{split}
\label{eq1}
\end{equation}
where $\hat{p}^{\dagger}$ ($\hat{q}^{\dagger}$) and $\hat{p}$ ($\hat{q}$) are creation and annihilation operators corresponding to the chain P (Q), $i$ ($j$) labels the site of an atom in the $x$ ($y$) direction, $\gamma=3.03\text{ eV}$ is the hopping integral, and $N_{x}$ is the number of unit cells in the $x$ direction. Using the similar method, the ZGNR Hamiltonian can be expressed by
\begin{equation}
\begin{split}
\hat{H}^{z}=&-\gamma\sum_{i=1}^{N_{x}}\sum_{j=1}^{N_{z}/2}\left(\hat{c}_{i,2j}^{\dagger}\hat{c}_{i,2j-1}+\hat{c}_{i,2j}^{\dagger}\hat{c}_{i,2j+1}\right.\\
&\left.+\hat{c}_{i,2j}^{\dagger}\hat{c}_{i+1,2j-1}\right)+\text{H.c.}.
\end{split}
\label{eq2}
\end{equation}

Since we have periodic boundary conditions in the $x$ direction, the Fourier transform can be made by
\begin{subequations}
	\begin{align}
		\hat{c}_{i,j}^{\dagger}&=\frac{1}{\sqrt{N_{x}}}\sum_{k_{x}\in\text{BZ}}e^{-ik_{x}x_{i}}\hat{c}_{k_{x},j}^{\dagger},\\
		\hat{c}_{i,j}&=\frac{1}{\sqrt{N_{x}}}\sum_{k_{x}\in\text{BZ}}e^{ik_{x}x_{i}}\hat{c}_{k_{x},j},
	\end{align}
\end{subequations}
where $k_{x}$ is the quasi-momentum of the $x$ direction, $x_{i}$ is the atomic position in the $x$ direction, and $\sum_{k_{x}\in\text{BZ}}$ is the summation over the Brillouin zone (BZ). Then Hamiltonians of AGNR and ZGNR can be written by
\begin{align}
\begin{split}
\hat{H}^{a}=&\sum_{k_{x}\in\text{BZ}}\sum_{j=1}^{N_{a}/2}\left[\gamma^{a}_{1}(\hat{p}_{k_{x},2j}^{\dagger}\hat{p}_{k_{x},2j-1}\right.\\
&+\hat{p}_{k_{x},2j}^{\dagger}\hat{p}_{k_{x},2j+1}+\hat{q}_{k_{x},2j-1}^{\dagger}\hat{q}_{k_{x},2j-2}\\
&+\hat{q}_{k_{x},2j-1}^{\dagger}\hat{q}_{k_{x},2j})+\gamma^{a}_{2}(\hat{p}_{k_{x},2j}^{\dagger}\hat{q}_{k_{x},2j}\\
&\left.+\hat{q}_{k_{x},2j-1}^{\dagger}\hat{p}_{k_{x},2j-1})\right]+\text{H.c.},
\end{split}\\
\begin{split}
\hat{H}^{z}=&\sum_{k_{x}\in\text{BZ}}\sum_{j=1}^{N_{z}/2}\left(\gamma^{z}\hat{c}_{k_{x},2j}^{\dagger}\hat{c}_{k_{x},2j-1}\right.\\
&\left.-\gamma\hat{c}_{k_{x},2j}^{\dagger}\hat{c}_{k_{x},2j+1}\right)+\text{H.c.},
\end{split}
\end{align}
where hopping integrals $\gamma^{a}_{1}=-\gamma e^{-i\frac{k_{x}a}{2\sqrt{3}}}$, $\gamma^{a}_{2}=-\gamma e^{i\frac{k_{x}a}{\sqrt{3}}}$ and $\gamma^{z}=-2\gamma\cos(k_{x}\frac{a}{2})$. We obtain the AGNR Hamiltonian in the form of $\hat{H}^{a}=\sum_{k_{x}}\mathbf{\Phi}_{k_{x}}^{a\dagger}\mathbf{H}^{a}_{k_{x}}\mathbf{\Phi}_{k_{x}}^{a}$ using bases of $\mathbf{\Phi}_{k_{x}}^{a}=(\hat{p}_{k_{x},1},\hat{q}_{k_{x},1},\hat{p}_{k_{x},2},\hat{q}_{k_{x},2},\cdots,\hat{p}_{k_{x},N_{a}},\hat{q}_{k_{x},N_{a}})^{\mathrm{T}}$, where the matrix notation of AGNR Hamiltonian $\mathbf{H}^{a}_{k_{x}}$ reads
\begin{equation}
\left(\begin{array}{ccccccccc}
	0 & \gamma_{2}^{a*} & \gamma_{1}^{a*}\\
	\gamma_{2}^{a} & 0 & 0 & \gamma_{1}^{a}\\
	\gamma_{1}^{a} & 0 & 0 & \gamma_{2}^{a} & \gamma_{1}^{a}\\
	& \gamma_{1}^{a*} & \gamma_{2}^{a*} & 0 & 0 & \gamma_{1}^{a*}\\
	&  & \ddots & \ddots & \ddots & \ddots & \ddots\\
	&  &  & \gamma_{1}^{a*} & 0 & 0 & \gamma_{2}^{a*} & \gamma_{1}^{a*}\\
	&  &  &  & \gamma_{1}^{a} & \gamma_{2}^{a} & 0 & 0 & \gamma_{1}^{a}\\
	&  &  &  &  & \gamma_{1}^{a} & 0 & 0 & \gamma_{2}^{a}\\
	&  &  &  &  &  & \gamma_{1}^{a*} & \gamma_{2}^{a*} & 0
\end{array}\right).
\label{armmatr}
\end{equation}
Likewise, we can write the ZGNR Hamiltonian in terms of $\mathbf{\Phi}_{k_{x}}^{z}=(\hat{c}_{k_{x},1},\hat{c}_{k_{x},2},\cdots,\hat{c}_{k_{x},N_{z}})^{\mathrm{T}}$, in the from of $\hat{H}_{z}=\sum_{k_{x}}\mathbf{\Phi}_{k_{x}}^{z\dagger}\mathbf{H}^{z}_{k_{x}}\mathbf{\Phi}_{k_{x}}^{z}$, where $\mathbf{H}_{k_{x}}^{z}$ is written as
\begin{equation}
\left(\begin{array}{cccccc}
0 & \gamma^{z}\\
\gamma^{z*} & 0 & -\gamma\\
& -\gamma^{*} & 0 & \gamma^{z}\\
&  & \ddots & \ddots & \ddots\\
&  &  & \gamma^{z*} & 0 & -\gamma\\
&  &  &  & -\gamma^{*} & 0
\end{array}\right).
\label{zgmatr}
\end{equation}
The Hamiltonian $\mathbf{H}^{a}_{k_{x}}$ is a $2N_{a}\times 2N_{a}$ matrix because the AGNR contains $2N_{a}$ sites per unit cell. For the ZGNRs, however, Hamiltonian $\mathbf{H}^{z}_{k_{x}}$ only has $N_{z}$ orthogonal eigenstates, because each unit cell is composed of $N_{z}$ sites.

We solve time-independent Schrödinger equation
\begin{equation}
\mathbf{H}(k_{x})\mathbf{S}_{i}(k_{x})=E_{i}(k_{x})\mathbf{S}_{i}(k_{x}),
\label{TISE}
\end{equation}
obtaining the eigenstates $\mathbf{S}_{i}(k_{x})$ and eigenvalues $E_{i}(k_{x})$ of the system. In Figs.~\ref{fig1}(c) and \ref{fig1}(d), we show the energy bands of AGNR and ZGNR in the case of $N_{a}=10$ and $N_{z}=10$, which are calculated by plugging Eqs.~(\ref{armmatr}) and (\ref{zgmatr}) into Eq.~(\ref{TISE}), respectively.

\subsection{\label{sec:method2}Coupling to an external laser field and numerical calculations}
In order to simulate the interaction of laser field and GNR, we use the time-dependent Schr{\"o}dinger equation
\begin{equation}
	i\frac{\partial}{\partial t}\mathbf{\Psi}(t)=\mathbf{H}(t)\mathbf{\Psi}(t),
\end{equation}
where $\mathbf{H}(t)$ is the matrix form of time-dependent Hamiltonian, which has been coupled to the external field $\mathbf{A}(t)$. 

Here, the vector potential $\mathbf{A}(t)=(0,A_{y})^{\text{T}}$ reads
\begin{align}\label{laser}
	 A_{y}(t)=\frac{E_{0}}{\omega}\sin^{2}{\left(\frac{\omega{t}}{2n_{cyc}}\right)}\sin(\omega t)
\end{align}
in the domain $0\leqslant t\leqslant n_{cyc} T$, where $E_{0}$ is the amplitude of the electric field, $\omega$ is the fundamental frequency, $T=2\pi/\omega$ is the time period, and $n_{cyc}$ is the total number of laser cycles.

For a tight-binding GNR, the driving fields shift the quasi-momentum $k_{x}$ like
\begin{equation}
	k_x\rightarrow k_{x}(t)={k_x}+{A}_{x}(t),
\end{equation}
because of the periodic boundary condition in the $x$ direction \cite{Graf1995}. The $y$ component of the drivers will induce phase factors on the hopping terms, which makes the replacement of
\begin{equation}
	\gamma_{ij}\rightarrow\gamma_{ij}e^{-iA_{y}(t)(y_{j}-y_{i})},
\end{equation}
where $\gamma_{ij}$ represents the hopping integral between site $i$ and site $j$ \cite{Graf1995}.

Thereby, according to the above statement, the matrix elements of the time-dependent AGNR Hamiltonian $\mathbf{H}^{a}(t)$ are expressed by
\begin{equation}
\begin{cases}
	\mathbf{H}_{i,i+1}^{a}[k_{x}(t)], & i=1,2,\ldots,2N_{a}-1\\
	\gamma_{y}^{a}(t)\cdot\mathbf{H}_{i,i+2}^{a}[k_{x}(t)], & i=1,2,\ldots,2N_{a}-2\\
	0, & \text{otherwise}
\end{cases}
\label{Hat}
\end{equation}
where the phase factor $\gamma_{y}^{a}(t)=e^{iaA_{y}(t)/2}$ and $\mathbf{H}_{ij}^{a}(k_{x})$ correspond to the elements in Eq.~(\ref{armmatr}). Similarly, the elements of ZGNR Hamiltonian $\mathbf{H}^{z}(t)$ read
\begin{equation}
\begin{cases}
	\gamma_{y,1}^{z}\cdot\mathbf{H}_{2i-1,2i}^{z}[k_{x}(t)], & i=1,2,\ldots,N_{z}/2\\
	\gamma_{y,1}^{z}\cdot\mathbf{H}_{2i,2i+1}^{z}[k_{x}(t)], & i=1,2,\ldots,N_{z}/2-1\\
	0, & \text{otherwise}
\end{cases}
\label{Hzt}
\end{equation}
where $\gamma_{y,1}^{z}(t)=e^{i\frac{aA_{y}}{2\sqrt{3}}}$, $\gamma_{y,2}^{z}(t)=e^{i\frac{aA_{y}}{\sqrt{3}}}$ and $\mathbf{H}_{ij}^{z}(k_{x})$ correspond to the elements in Eq.~(\ref{zgmatr}). It is important to note that only the upper triangular elements of $\mathbf{H}(t)$ are given in Eqs.~(\ref{Hat}) and (\ref{Hzt}), and one can easily obtain the values of lower triangular elements by using $\mathbf{H}_{ji}=\left(\mathbf{H}_{ij}\right)^{*}$.

In this simulation, the initial states $\mathbf{\Psi}_{nk_{x}}(0)$ are determined by the half lowest eigenstates of the Hamiltonian. The wave functions are numerically propagated with the Crank-Nicolson method \cite{Crank1947}:
\begin{align}
	\mathbf{\Psi}(t+\Delta t)\simeq\left[{\mathbf{I}-\frac{\mathbf{H}(t)\Delta t}{2i}}\right]^{-1}\left[{\mathbf{I}+\frac{\mathbf{H}(t)\Delta t}{2i}}\right]\mathbf{\Psi}(t),
\end{align}
where $\mathbf{I}$ is the identity matrix, and $\Delta t$ is the discrete time step for temporal evolution. Then we can calculate the value of induced electric current as
\begin{equation}
	 J_{\alpha}(t)=-\sum^{N/2}_{n=1}\sum_{k_{x}\in\text{BZ}}\mathbf{\Psi}^{\dagger}_{nk_{x}}(t)\left(\frac{\partial\mathbf{H}}{\partial A_{\alpha}}\right)\mathbf{\Psi}_{nk_{x}}(t),
\end{equation}
where $n$ labels the state, $N$ is the size of the Hamiltonian matrix, $\alpha=x$, $y$, and  $\mathbf{\Psi}^{\dagger}_{nk_{x}}=\left(\mathbf{\Psi}^{*}_{nk_{x}}\right)^{\text{T}}$. Ultimately, the harmonic spectrum along $\alpha$ direction can be evaluated from the current by
\begin{equation}
	S_{\alpha}(\omega)=\mid\text{FT}[\frac{d}{d t} J_{\alpha}(t)]\mid^2.
\end{equation}
Note that the derivative of current is multiplied by a Blackman window before the Fourier transform.

\section{\label{sec:selection_rules}Optical selection rules in GNRs}

\begin{figure}
	\centering
	\includegraphics[width=\columnwidth]{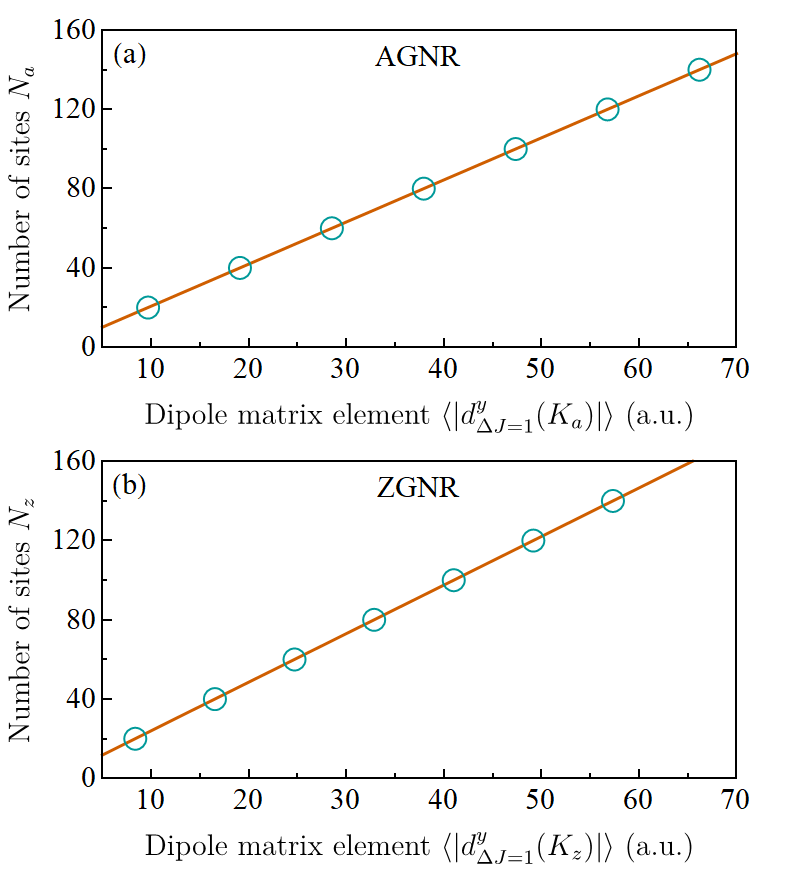}
	\caption{Dipole matrix elements $\langle \mid d_{\Delta J=1}^{y}(K_{l})\mid\rangle$ with respect to the ribbon widths $N_{l}$ for (a) AGNRs $(l=a)$ and (b) ZGNRs $(l=z)$. The open circles are data points, and orange lines are linear fit of data points.}
	\label{fig2}
\end{figure}

Since the selection rules of GNR-subbands \cite{MingFa2000,Hsu2007,Chung2011,Sasakiprb2011,Saroka2017} are crucial for understanding the size-dependent HHG spectra in Sec.~\ref{sec:III}, it is necessary for us to clarify them in this section.

The dipole matrix element between subbands $J_{1}(s)$ and $J_{2}(s')$ reads
\begin{equation}
	 d_{J_1s,J_2s'}^{\alpha}=\langle \phi^{s}_{J_1}\mid \hat{r}_{\alpha}\mid\phi^{s'}_{J_2} \rangle=-i\frac{\langle\phi^{s}_{J_1}\mid\hat{p}_{\alpha}\mid\phi^{s'}_{J_2}\rangle}{E_{J_1}-E_{J_2}},
	\label{dd}
\end{equation}
where $\hat{r}_{\alpha}$ is position operator along $\alpha$ direction, $\hat{p}_{\alpha}$ is the momentum operator along $\alpha$ direction, $s$ and $s'$ are the
band type indices, $J_{1}$ and $J_{2}$ are the subband numbers, and $E_{J}$ is the band dipersion of
subband $J$. These indices have been introduced in Sec.~\ref{sec:tb}. In what follows, only the $y$ component of the dipole matrix element $d_{J_{1}s,J_{2}s'}^{y}$ is discussed, because we mainly focus on the optical transition while the GNR is driven by $y$-polarized light.
In Eq.~(\ref{dmatelem}), we give a example of dipole matrix (modulus) of $10$-ZGNR at Dirac point $K_{z}$,
\begin{eqnarray}
&&\mid d^{y}_{J_{1}s,J_{2}s'}(K_{z})\mid= \nonumber \\
&&\left(
\scalemath{0.825}{
\begin{array}{cccccccccc}
0. & 3.99 & 0. & 0.32 & 0. & 0.084 & 0. & 0.03 & 0. & 0.33\\
3.99 & 0. & 4.30 & 0. & 0.40 & 0. & 0.113 & 0. & 0.297 & 0.\\
0. & 4.30 & 0. & 4.39 & 0. & 0.43 & 0. & 0.213 & 0. & 0.03\\
0.32 & 0. & 4.39 & 0. & 4.42 & 0. & 0.103 & 0. & 0.114 & 0.\\
0. & 0.40 & 0. & 4.42 & 0. & 4.09 & 0. & 0.43 & 0. & 0.084\\
0.084 & 0. & 0.43 & 0. & 4.09 & 0. & 4.42 & 0. & 0.40 & 0.\\
0. & 0.113 & 0. & 0.103 & 0. & 4.42 & 0. & 4.39 & 0. & 0.32\\
0.03 & 0. & 0.213 & 0. & 0.43 & 0. & 4.39 & 0. & 4.30 & 0.\\
0. & 0.297 & 0. & 0.114 & 0. & 0.40 & 0. & 4.30 & 0. & 3.99\\
0.33 & 0. & 0.03 & 0. & 0.084 & 0. & 0.32 & 0. & 3.99 & 0.
\end{array}
}
\right), \nonumber \\
\label{dmatelem}
\end{eqnarray}
in the basis $ (|5v\rangle$, $|4v\rangle$, \ldots $|1v\rangle$, $|1c\rangle$, \ldots $|5c\rangle)^{\rm T}$. We can see some zero matrix elements which result in the interesting selection rule in ZGNRs.

Firstly, let us see the selection rule in AGNRs. According to the analytical derivations in Refs.~\cite{Chung2011,Sasakiprb2011,Saroka2017} and our numerical calculations, the dipole matrix elements of intraband from the same group, i.e., $s'=s$, and $(J_{1},J_{2})\in \left\{J\right\}$ or $\left\{J'\right\}$, possess the following feature:
\begin{equation}
	\begin{cases}
		d_{J_{1}s,J_{2}s}^{y}=0, & \Delta J=J_{2}-J_{1}\in\text{even}\\
		d_{J_{1}s,J_{2}s}^{y}\neq 0, & \Delta J=J_{2}-J_{1}\in\text{odd}.
	\end{cases}
\label{dipolec1}
\end{equation}
For the interband transition ($s\neq s'$), dipole matrix elements from the same group show
\begin{equation}
	\begin{cases}
		d_{J_{1}s,J_{2}s'}^{y}\neq 0, & \Delta J=J_{2}-J_{1}\in\text{even}\\
		d_{J_{1}s,J_{2}s'}^{y}= 0, & \Delta J=J_{2}-J_{1}\in\text{odd}.
	\end{cases}
\label{dipolecinter}
\end{equation}
That is to say the intraband (interband) transition between those two subbands is forbidden whenever the difference in corresponding indices ($\Delta J$) is an even (odd) number. Then let us move to another case, considering the transition between the subbands in different groups, that is $J_{1}\in \left\{J\right\}$ and $J'_{2}\in\left\{J'\right\}$. The intraband ($s=s'$) dipole moments are shown as
\begin{align}
    \begin{cases}
		d_{J_{1}s,J'_{2}s}^{y}\neq0, & \Delta J=J'_{2}-J_{1}\in\text{even}\\
		d_{J_{1}s,J'_{2}s}^{y}=0, & \Delta J=J'_{2}-J_{1}\in\text{odd}.
	\end{cases}
\label{dipolec2}
\end{align}
For the interband transition ($s\neq s'$), the situation is just opposite to Eq.~(\ref{dipolec2}), that is, the optical transition is forbidden whenever the index difference $\Delta J$ is even, shown as
\begin{align}
    \begin{cases}
		d_{J_{1}s,J'_{2}s'}^{y}=0, & \Delta J=J'_{2}-J_{1}\in\text{even}\\
		d_{J_{1}s,J'_{2}s'}^{y}\neq0, & \Delta J=J'_{2}-J_{1}\in\text{odd}.
	\end{cases}
\label{dipmtinter2}
\end{align}

Secondly, we consider the selection rule of ZGNR which is simpler than that of AGNR, because the ZGNR only has one group of subbands. Based on the similar calculation and analysis, we find that the features of dipole matrix elements for intraband and interband transitions are just same as which has been shown in Eqs.~(\ref{dipolec1}) and (\ref{dipolecinter}). The intraband transition is allowed when the index difference $\Delta J$ is 
odd, whereas the interband transition is allowed when the index difference $\Delta J$ is even.

Focusing on the intraband transitions near the Dirac points, we compute the dipole matrix elements  $d^{y}_{J,J+1}$ for AGNR and ZGNR of nearest-neighbor subbands at Dirac points using Eq.~(\ref{dd}). It shows an interesting characteristic that the moduli of $d^{y}_{J,J+1}(K)$ are almost the same with different band numbers $J$, for example, $d^{y}_{\Delta J=1}(K_{z})$ in Eq.~(\ref{dmatelem}). So we plot $\langle \mid d_{\Delta J=1}^{y}(K_{a})\mid\rangle$ and $\langle \mid d_{\Delta J=1}^{y}(K_{z})\mid\rangle$, the average values of dipole matrix elements of nearest-neighbor subbands at Dirac points, in Figs.~\ref{fig2}(a) and \ref{fig2}(b), respectively. For AGNR, the results of matrix elements from the second group $\left\{J'\right\}$ are not presented in Fig.~\ref{fig2}(a). We can see clearly that the dipole matrix elements $\langle \mid d_{\Delta J=1}^{y}\mid\rangle$, for both AGNR and ZGNR, vary linearly with the number of sites:
\begin{subequations}
	\begin{align}
		\label{d1}
		\langle \mid d_{\Delta J=1}^{y}(K_{a})\mid\rangle=0.47N_{a}+0.26,\\
		\label{d2}
		\langle \mid d_{\Delta J=1}^{y}(K_{z})\mid\rangle=0.41N_{z}+0.20.
	\end{align}
\end{subequations}
This indicates that the coupling between nearest-neighbor subbands becomes stronger while increasing the ribbon width.

\section{Rise time depending on total number of energy levels and coupling strength in the equi-spaced model}\label{appenrise}
\begin{figure}
	\centering
	\includegraphics[width=\columnwidth]{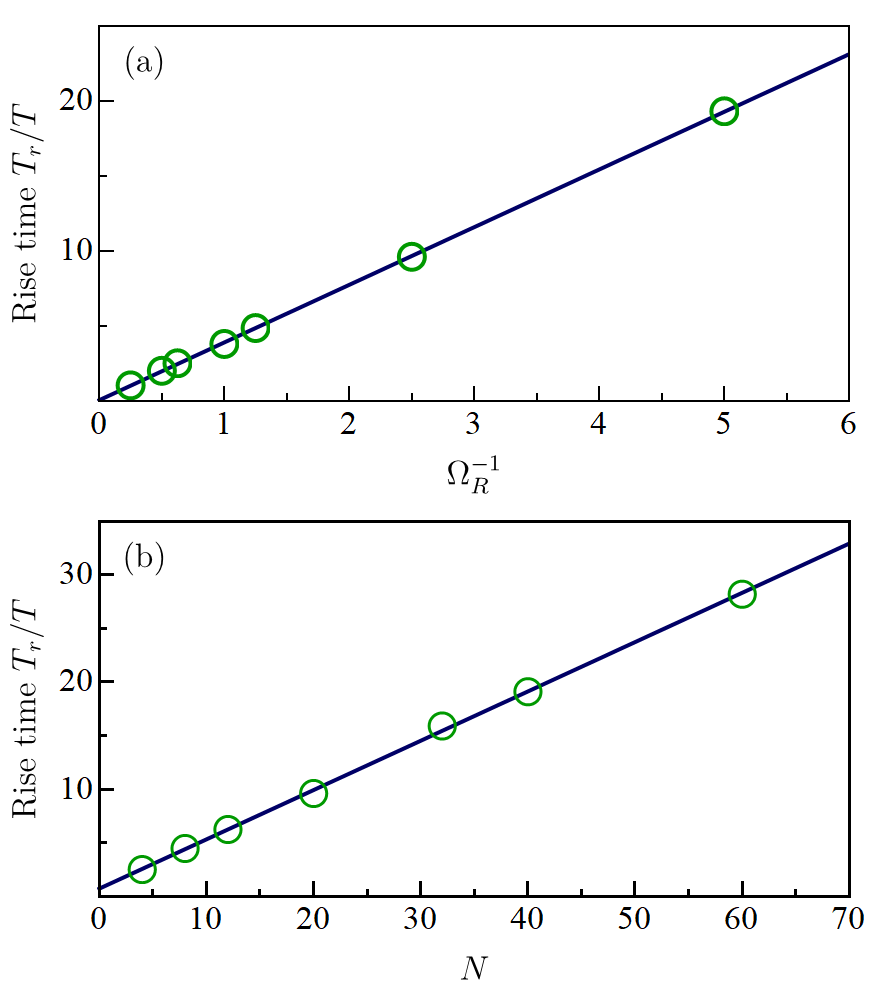}
	\caption{(a) Linear rise time as a function of $\Omega_{R}^{-1}$ for $N=20$. The dipole matrix elements are same as in Eq.~(\ref{dipmatelem}). Laser frequency is set to $\omega=\Delta_{\rm sg}=1$ a.u.. (b) Linear rise time as a function of $N$. We fix field strength $E_{0}$ at $0.4$ a.u., and thus $\Omega_{R}=E_{0}\cdot d_{i,i+1}=0.4$ a.u.. Laser frequency is set to $\omega=\Delta_{\rm sg}=1$ a.u.. Green open circles represent data points, and the blue lines are linear fitting results for data points.}
	\label{risetime}
\end{figure}
We know that in equi-spaced model the cascade resonance occurs when frequency condition is satisfied no matter how weak the electric field is. The electrons are excited to conduction levels in a cascade way from lower to higher. We can foresee that the weaker the field strength and the more the energy levels, the longer it takes for the electrons to be excited to the highest energy level. Therefore, the rise time should depend on the total number of energy levels and the coupling strength.
Figures~\ref{risetime}(d) and \ref{risetime}(e) clearly show that the rise time exhibits a linear dependence of $\Omega^{-1}_{R}$ and $N$. We can describe it by
\begin{align}
    T_{r} \propto N/\Omega_{R}.
\end{align}
From the perspective of HHG, this relation indicates that the manipulation of the harmonic cutoff using laser duration is easier to achieve in a large-size nanoribbon material.

\section{HHG in GNRs by applying driving fields along the ribbon direction}\label{ribbonHHG}
\begin{figure}
	\centering
	\includegraphics[width=\columnwidth]{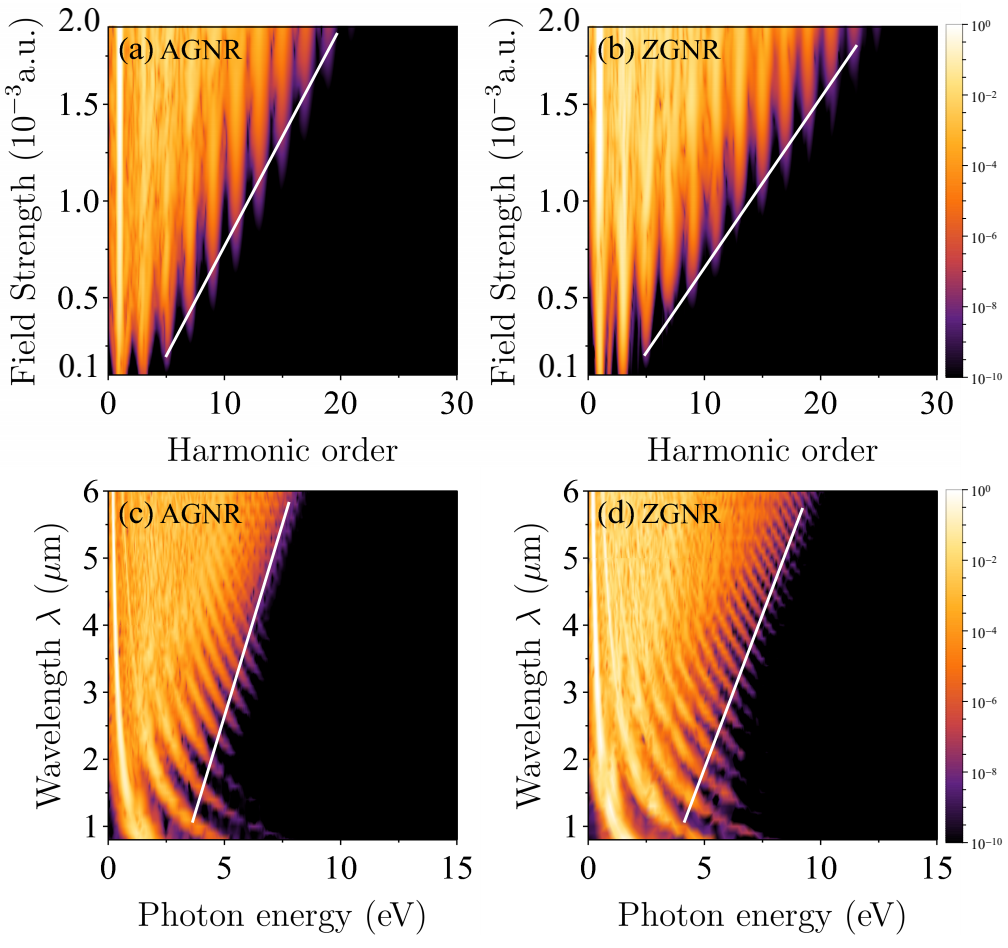}
	\caption{(a)--(b) Harmonic spectra vs field strength in $40$-AGNR and $40$-ZGNR, respectively, when the field is polarized along the ribbon direction. The laser frequency is fixed at $\omega_{0}=0.0152\text{ a.u. }(\lambda=3\;\mu\text{m})$. (c)--(d) Harmonic spectra vs wavelength in $40$-AGNR and $40$-ZGNR, respectively. The field strength is $E_{0}=0.0012$ a.u. ($I_{0}=5.04\times 10^{10}$ W/cm$^2$).}
	\label{figA1}
\end{figure}
We study the harmonic generation in GNRs by applying the driver field to the ribbon direction. Figures~\ref{figA1}(a) and \ref{figA1}(b) respectively show the harmonic spectra for $40$-AGNR and $40$-ZGNR as a function of the field strength, and the wavelength is $3$ $\mu$m. It is found that the cutoffs of harmonic spectra for both AGNR and ZGNR scale linearly with the field strength, see the white lines. It is because the dispersion of subbands of GNRs near the Fermi level is almost linear with $k_x$. In Figs.~\ref{figA1}(c) and \ref{figA1}(d), we also see the linear dependence between the cutoff energy and the laser wavelength, which agrees well with the general behavior of HHG in bulk solids. The reason is also attributed to the linear band dispersion.

\section{HHG in GNRs with onsite potential}\label{sec:gappedZGNR}
\begin{figure}
	\centering
	\includegraphics[width=\columnwidth]{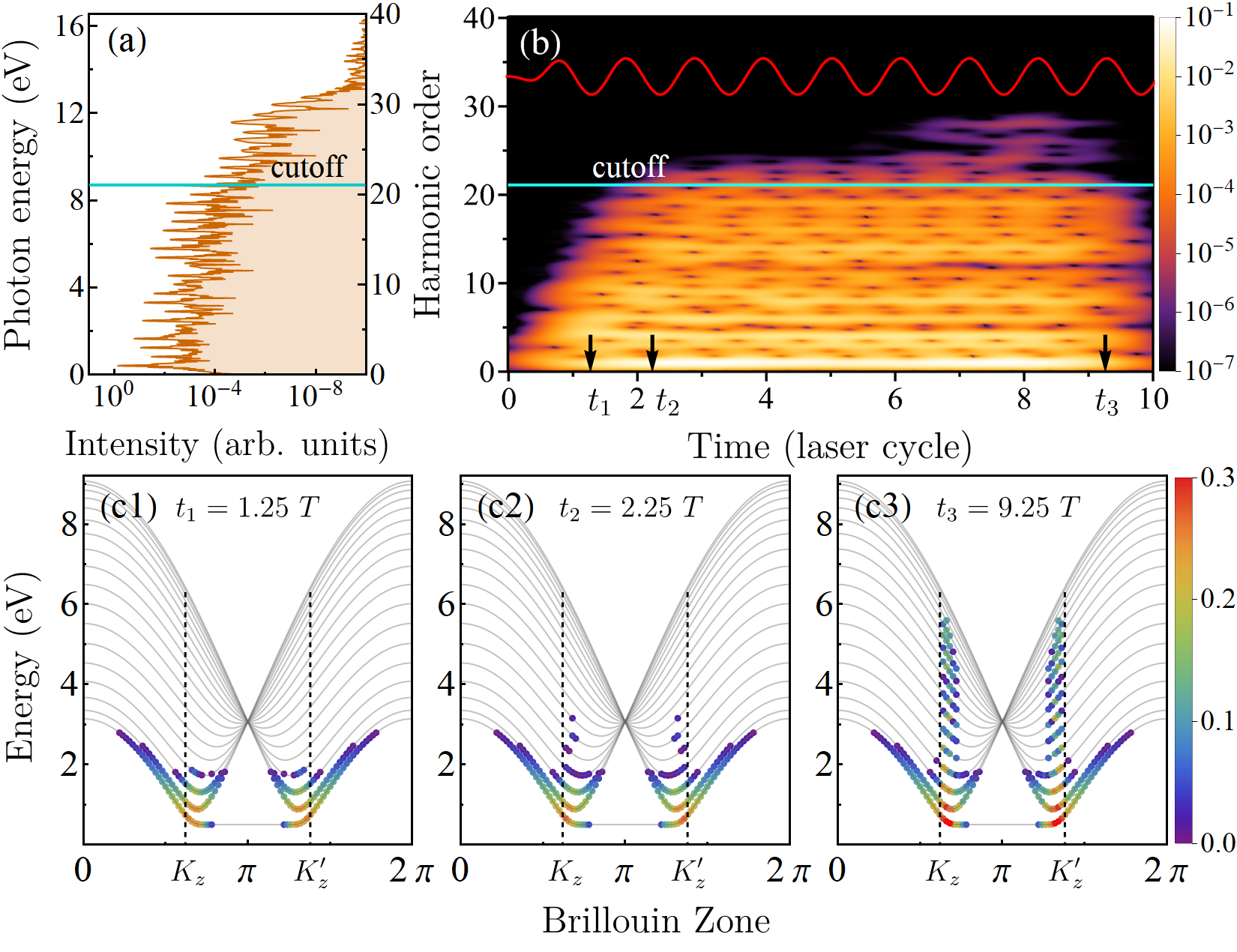}
	\caption{(a) Harmonic spectrum for a 36-ZGNR with $E_{\rm g}=1.0$ eV. The field strength is $E_{0}=0.0012$ a.u. ($I_{0}=5.04\times 10^{10}$ W/cm$^2$), and the field frequency matches subband gap $\omega=0.0152$ a.u. ($\lambda=3\;\mu$m). (b) Time-frequency distribution of the harmonic emission. The red curve is the vector potential of laser field. The cyan line indicates the cutoff frequency. (c1)--(c3) Electron dynamics in the conduction subbands for the times $t_{i}$ marked in (b).}
	\label{gappedZGNR}
\end{figure}

In principle, we can open a fundamental gap in GNRs by adding staggered onsite potential on the two nearest-neighbor sites respectively.
Figure~\ref{gappedZGNR} shows the HHG spectrum and conduction dynamics in $36$-ZGNR with a $1.0$ eV gap. Since a large fundamental gap between valence and conduction bands, the excitation rate decreases, and fewer electrons are excited to the first conduction band in Figs.~\ref{gappedZGNR}(c1)---(c3). However, one can still observe the cascade resonance, and the electrons still accumulate near the Dirac points although the total population becomes lower in comparison to that of gapless $36$-ZGNR. Correspondingly, the intensity of the HHG spectrum in the Figs.~\ref{gappedZGNR}(a) and \ref{gappedZGNR}(b) is lower than that in the Figs.~\ref{zig40edyna}(a) and \ref{gappedZGNR}(b).

\bibliographystyle{apsrev4-1_title}
\bibliography{hhg}

\end{document}